\documentclass[11pt]{article}
\usepackage[pdftex]{graphicx}
\usepackage{hyperref}
\usepackage{float} 
\usepackage{multirow}
\hypersetup{
	colorlinks   = true, 
	urlcolor     = blue, 
	linkcolor    = blue, 
	citecolor   = red 
}

\usepackage{cite}
\usepackage{multirow}
\usepackage{amsmath,amsfonts,amssymb,amsthm,nccmath,latexsym,mathtools}
\usepackage{xcolor}
\usepackage{amsmath}

\usepackage{tikz,lipsum,lmodern} 
\usepackage[]{tcolorbox}
\definecolor{light-gray}{gray}{0.95}

\usepackage{dsfont}
\usepackage{cancel} 
\usepackage{ amssymb }
\usepackage{footnote}

\def\o{\omega}

\def\no{\nonumber}
\def\a{\alpha}
\def\b{\beta}

\def\d{\delta}

\def\p{\partial}

\def\na{\nabla}
\def\tna{\tilde{\nabla}}
\def\T{\Theta}

\def\k{\kappa}

\def\t{\tilde}

\def\be{\begin{equation}}
\def\ee{\end{equation}}
\def\ba{\begin{align}}
\def\ea{\end{align}}

\def\mg{\sqrt{-g}}

\def\tmg{\sqrt{-\tilde g}}
\def\K{\mathcal{K}}
\def\tK{\tilde{\mathcal{K}}}
\def\H{\mathcal{H}}
\def\B{\mathcal{B}}

\title{Boundary terms and Brown-York quasi-local parameters for scalar-tensor theory: a study on both timelike and null hypersurfaces}

\author{Krishnakanta Bhattacharya\footnote{krish.phys@gmail.com} \ and \ Kazuharu Bamba\footnote{bamba@sss.fukushima-u.ac.jp}\\
Faculty of Symbiotic Systems Science, Fukushima University, Fukushima 960-1296, Japan.}

\date{\today}
\begin{document}
\maketitle

\begin{abstract}
Boundary term and Brown-York (BY) formalism, which is based on the Hamilton-Jacobi principle, are complimentary of each other as the gravitational actions are not, usually, well-posed. In scalar-tensor theory, which is an important alternative to GR, it has been shown that this complementarity becomes even more crucial in establishing the equivalence of the BY quasi-local parameters in the two frames which are conformally connected. Furthermore, Brown-York tensor and the corresponding quasi-local parameters are important from two important yet different aspects of gravitational theories: black hole thermodynamics and fluid-gravity correspondence. The investigation suggests that while the two frames are equivalent from the thermodynamic viewpoints, they are not equivalent from the perspective of fluid-gravity analogy or the membrane paradigm. In addition, the null boundary term and null Brown-York formalism are the recent developments (so far obtained only for GR) which is non-trivial owing to the degeneracy of the null surface. In the present analysis these are extended for scalar-tensor theory. The present analysis also suggests that, regarding the equivalence (or inequivalence) of the two frame, the null formalism draws the same inferences as of the timelike case, which in turn establishes the consistency of the newly developed null Brown-York formalism.

\end{abstract}
\section{Introduction}
Several theoretical analysis \cite{Donoghue:1995cz, Donoghue:2012zc, Burgess:2003jk} and observational data \cite{Riess:2001gk, Riess:2004nr, KNOP, Perlmutter:1998np,  Tonry:2003zg, Barris:2003dq, Perlmutter:1997zf, Riess:1998cb, RIESS4} suggests that Einstein's general relativity (GR) is not the ultimate theory of gravitation. As a result, the interests on modified theories of gravity has surged over time. Among several modified theories of gravity of special interests, the scalar-tensor (ST) theory is, probably, the most popular for several reasons. The scalar-tensor theory is favoured by the string theory \cite{Damour:2002mi} as the spin-2 graviton has a spin-0 partner known as dilaton. Therefore, as per the string theory, the actual theory of gravity should be a scalar-tensor gravity. Furthermore, the higher curvature gravity ($f(R)$ theory) can be studied equivalently as a scalar-tensor theory \cite{Nojiri:2006ri}. As a result, the scalar-tensor theory is considered as one of the most important both from the theoretical and observational viewpoints \cite{Faraoni:2004pi,FUJII,Copeland:2006wr,Elizalde:2008yf,Nojiri:2010wj,Nojiri:2017ncd,Hinterbichler:2011tt,Amendola:2015ksp,Clifton:2011jh,Nordtvedt:1970uv,Billyard:1998kg,Barrow:1996kc,Mimoso:1995ge,Barrow:1994nx,Matsuda:1972zp,Romero:1992xx,Romero:1992bu,Romero:1992ci,Paiva:1993bv,Paiva:1993qa,Scheel:1994yn,Anchordoqui:1997du,Banerjee:1996iy,Faraoni:1999yp,Bertotti:2003rm,Bettoni:2016mij,Baker:2017hug,Sakstein:2017xjx}. Furthermore, the scalar-tensor theory is described in the two frames, which are conformally connected. From the naive perspective of quantum field theory, one expects that the physical parameters should be equivalent under the conformal transformation, which is basically a field re-parametrization. Therefore, scalar-tensor theory provide a perfect toy model to verify such argument. The equivalence/ inequivalence of the two frames has been the subject of intense research for a long time and it is still considered as a matter for debate \cite{Faraoni:1999hp, ALAL,Faraoni:1998qx, Faraoni:2010yi, Faraoni:2006fx, Saltas:2010ga, Capozziello:2010sc, Padilla:2012ze, Koga:1998un, Jacobson:1993pf, Kang:1996rj, Deser:2006gt, Dehghani:2006xt, Sheykhi:2009vc,Steinwachs:2011zs, Kamenshchik:2014waa, Banerjee:2016lco, Pandey:2016unk, Ruf:2017xon,Karam:2017zno,Bahamonde:2017kbs,Karam:2018squ,Bhattacharya:2017pqc,Bhattacharya:2018xlq,Bhattacharya:2020wdl,Bhattacharya:2020jgk,Dey:2021rke,Bhattacharya:2021lgk,Bhattacharya:2022mnb,Seenivasan:2023utj}.

For all the importances of the ST gravity stated above, the theory should be studied in a comprehensive manner. The action of the ST gravity is not well-posed. Since the action (in each frames) of the ST gravity contains the first-order as well as the second-order derivative of the metric tensor, one has to fix both metric as well as its first order derivative on the boundary. As a result, the principle of least action becomes ill-defined. To resolve this issue, there are two choices, which we have discussed subsequently in detail. One of the choices, which is a non-covariant formalism, has been explored in earlier works \cite{Bhattacharya:2017pqc,Bhattacharya:2020jgk} (also see the review \cite{Bhattacharya:2022mnb}). There remains another choice \textit{i.e.} addition of a suitable boundary term which negates the problematic terms, which causes the action principle ill-defined. As it has been explored in general relativity, the boundary term is not unique \cite{Charap:1982kn}. There can be several boundary terms which can be used in order to obtain the well-posed action principle. On the contrary, it has been argued recently in literature that we should not pre-impose boundary term with the action. Instead, the action principle should tell us what boundary term is required to be added so that when the action is well-posed, the number of components which are required to be fixed on the boundary should be equal to the number of true degrees of freedom in the theory \cite{Padmanabhan:2014lwa,Parattu:2015gga,Parattu:2016trq,Chakraborty:2017zep,Chakraborty:2018dvi}. Although, the boundary term required for the well-posed action (on timelike/spacelike surface) is known for ST gravity, an analysis in such spirit is still missing, where the boundary term is decided by the action principle itself. Furthermore, there is another major challenge. The formalism of obtaining the boundary term from the action principle itself is a foliation-dependent formalism. As a result, the usual approach for a timelike (or spacelike) surface, will not be valid for a null surface, which is degenerate and possess several non-trivial traits as compared to the timelike (or spacelike) counterparts. Even in general relativity (GR), the proper boundary term for a null surface has been obtained very recently \cite{Parattu:2015gga,Parattu:2016trq,Lehner:2016vdi,Hopfmuller:2016scf,Oliveri:2019gvm,Aghapour:2018icu,Chandrasekaran:2020wwn} (which has later been extended for the Lanczos-Lovelock gravity \cite{Chakraborty:2018dvi}). Whereas, the boundary term required for the actions of ST gravity is not yet defined for the null surfaces. Once the boundary terms are known for timelike/spacelike and the null surfaces, one can obtain the Brown-York tensor and the corresponding quasi-local parameters, which are important both from the perspectives of black hole thermodynamics as well as membrane-paradigm (or fluid-gravity correspondence) \cite{Brown:1992br,Brown:2000dz}. Although from the previous analysis \cite{Creighton:1995au,Bose:1998yp,Cote:2019fkf} it is known how the Brown-York energy and mass are conformally related, the connections of other parameters are not yet studied. More importantly, for the null surface, the Brown-York tensor and the quasi-local parameters are not yet obtained in the two frames. So far, only in GR \cite{Jafari:2019bpw,Chandrasekaran:2021hxc}, the BY tensor and the quasi-local parameters are obtained for a null surface owing to its non-trivial properties. As a result it is not known how the null quasi-local parameters are conformally connected and whether it yields the same connection as of the timelike case.

  In the present analysis, we address all these issues which are described above and obtain all-around understanding in this context. We have presented subsequent analysis separately for the timelike/spacelike surface and for the null surface, where we obtain the boundary term from the action principle itself and, thereafter, the BY tensor and the corresponding quasi-local parameters. We also compare how the quasi-local parameters are connected in the two frames and, thereby, what implication it makes for black hole thermodynamics and membrane paradigm under the conformal transformation. The paper is organized as follows: In the following section (section \ref{sect2}) we provide a brief review of scalar-tensor gravity in the two frames, where we show how the two actions are conformally connected and also discuss the equivalence/ inequivalence of the two frames at the action level. In section \ref{sectimelike}, we have presented the analysis of obtaining the boundary term and quasi-local parameters for the timelike/spacelike surface. We also obtain how the parameters in the two frames are connected. In section \ref{secnull}, we obtain the same analysis for a more non-trivial case \textit{i.e.} for the null surface. In section \ref{seccompare}, we have compared the analysis in timelike/spacelike surface with that of the null surface. Finally in section \ref{secconcl}, we provide the conclusions of our analysis. 
 \vskip 2mm
 \textit{Units and notations:} Here we have adopted the geometrized units and have set $\hbar=c=G=1$. Furthermore, quantities with a tilde overhead (such as $\t A$) will correspond to those of the Einstein frame and quantities without a tilde (such as $A$) will correspond to those of the Jordan frame.

\section{Scalar-tensor gravity: A brief review} \label{sect2}
The scalar-tensor (ST) gravity is described in the two frames. The original frame is known as the Jordan frame where we have a non-minimal coupling of scalar field and the Ricci-scalar in the action. As a result, the gravity is mediated not only by the metric tensor, but also the scalar field $\phi$. The action in the Jordan frame is given as
\begin{eqnarray}
&&\mathcal{A}=\int_\nu d^4x\sqrt{-g}L =\int_\nu d^4x\sqrt{-g}\Big[ \frac{1}{16\pi}\Big(\phi R
-\frac{\omega (\phi)}{\phi}g^{ab}\nabla_a\phi \nabla_b\phi -V(\phi)\Big)\Big]~.
\label{SJ}
\end{eqnarray}
For a generic ST gravity, which we have considered in our analysis, $\o(\phi)$ is considered as a function of $\phi$. When $\omega$ is a constant, it boils down to the Brans-Dicke theory \cite{Brans:1961sx}. Now, the above action \eqref{SJ} has an Einstein frame representation, where the non-minimal coupling is no longer present. From Jordan frame, one can arrive the Einstein frame via the following transformation relations: (i) a conformal transformation in the metric along with (ii) a rescaling in the scalar field $\phi$. The transformation relations are provided as

\begin{align}
g_{ab}\rightarrow\tilde{g}_{ab}=\Omega^2g_{ab},\ \ \ \ \ \ \ \ \Omega=\sqrt{\phi}~,
\label{GAB}
\end{align}
and 
\begin{align}
 \phi\rightarrow\tilde{\phi}\,\ {\textrm{with}}\,\ d\tilde{\phi}=\sqrt{\frac{2\omega(\phi)+3}{16\pi}}\frac{d\phi}{\phi}~.
\label{PHI}
\end{align}
With the above transformations \eqref{GAB} and \eqref{PHI}, the above action \eqref{SJ} can be written in the Einstein frame as
\begin{eqnarray}
&&\tilde{\mathcal{A}}=\int_\nu d^4x\sqrt{-\tilde{g}}\tilde{L}=\int_\nu d^4x\sqrt{-\tilde{g}}\Big[\frac{\tilde{R}}{16\pi}-\frac{1}{2}\tilde{g}^{ab}\tilde{\nabla}_a\tilde{\phi}\tilde{\nabla}_b\tilde{\phi}-U(\tilde{\phi})\Big]~,
\label{SE}
\end{eqnarray}
where $U(\tilde{\phi}) = V(\phi)/(16\pi\phi^2)$. As we have mentioned earlier, in the Einstein frame (which is described by the action \eqref{SE}) the non-minimal coupling is no longer present.
Note, that the two actions \eqref{SJ} and \eqref{SE} are not exactly equivalent. In fact, the exact relation of the two actions are provided as follows:
\begin{align}
\t{\mathcal{A}}=\mathcal{A}-\frac{3}{16\pi}\int_{\nu}\mg\square\phi ~d^4x~.
\label{ACTUAL}
\end{align}
 The origin of the extra term ($\square\phi$) can be traced back from the conformal connection of the Ricci-scalar, which is provided as
\begin{align}
\tilde{R}=\frac{1}{\phi}\Big[R+\frac{3}{2\phi^2}(\nabla_i\phi)(\nabla^i\phi)-\frac{3}{\phi}\square\phi\Big]~. \label{RRCONF}
\end{align}

 However, since the last term (\textit{i.e.} the $\square\phi$ term) in \eqref{ACTUAL} is a total derivative term, it has been disregarded over the ages as it does not alter the dynamics of the system. Recently, it has been proved that this term plays a crucial role in the context of the equivalence of the two frames \cite{Bhattacharya:2017pqc,Bhattacharya:2018xlq,Bhattacharya:2020jgk,Bhattacharya:2022mnb}. In the present case, we shall see that the above relation \eqref{ACTUAL}  will help us again in establishing the equivalence in the two frames (see BY formalism in null surface). 
 
 We know that Brown-York formalism  of obtaining the surface stress tensor is based on the Hamilton-Jacobi (HJ) Principle \cite{Brown:1992br,Brown:2000dz}. But, in the case of gravitational action (in GR as well as in modified theories of gravity such as scalar-tensor gravity), it is a bit non-trivial, and should be explained properly. Let $L\equiv L(q^A(t),\dot q^A(t), t)$ be the classical Lagrangian of a dynamical system, then the variation of the action ($\mathcal{A}_{(cl)}$) yields
\begin{align}
\d\mathcal{A}_{(cl)}=\d\int_{t_1}^{t_2}L~dt=\int_{t_1}^{t_2}\Big[\frac{\p L}{\p q^A}-\frac{d}{dt}\Big(\frac{\p L}{\p\dot{q}^A}\Big)\Big]\d q^A~dt+\frac{\p L}{\p \dot{q}^A}\d q^A\Big|_{t_1}^{t_2}~. \label{VARCLASSIC}
\end{align} 
Thus, extremizing the action and fixing the boundary conditions in such a way that $q^A(t)$ are fixed at the end points (boundary), one obtains the Euler-Lagrange's equations. Moreover, the above equation \eqref{VARCLASSIC} also implies that if we restrict the variation of the action among the solutions of the dynamical system (\textit{i.e.} onshell variation), we obtain the expression of energy ($H$) and momentum ($P_A$) at the final boundary ($\lambda_2\equiv\lambda_2(q^A(t_2), t_2)$) from the Hamilton-Jacobi principle as follows:
\begin{align}
\frac{\p\mathcal{A}_{(cl)}}{\p q^A(t_2)}=\frac{\p L}{\p \dot{q}^A}\Big|_{\lambda_2}=P_A|_{\lambda_2}~.
\end{align}
and 
\begin{align}
\frac{\p \mathcal{A}_{(cl)}}{\p t}\Big|_{\lambda_2}=\Big(\frac{d \mathcal{A}_{(cl)}}{d t}-\frac{\p \mathcal{A}_{(cl)}}{\p q^A}\dot{q}^A\Big)\Big|_{\lambda_2}=\Big(L-P_A\dot{q}^A\Big)\Big|_{\lambda_2}=H|_{\lambda_2}~.
\end{align}

Based on the above HJ principle, Brown-York stress-tensor (or the surface stress tensor) is defined in GR. However, for the above definitions to work well, the pre-requisite is that the action-principle should be well-posed beforehand. Since the gravitational action contains the second order derivative of the fields (metric tensor and the scalar field), the above discussion, unfortunately, cannot be applied in a straightforward manner. For that, let us consider another example. Let us consider a pair of Lagrangians $L_1(q^A,\p q^A)$ and $L_2(q^A,\p q^A,\p^2 q^A)$ where $L_2$ has the specific form 
\begin{align}
L_2(q^A,\p q^A,\p^2 q^A)=L_1(q^A,\p q^A)-\p_i f^i(q^A,\p q^A),
\end{align}
here $i=\{0,1,2,3\}$ represents the spacetime indices and $A=\{1,2,3.....N\}$ labels all the generalized coordinates and the corresponding conjugate momentum. Although the Lagrangian $L_2$ contains second order derivative of $q^A$, the equation of motion is still second-order (and the same as one provided by $L_1$) as the second order derivative of $q^A$ appears in terms of a total derivative (surface term) in the Lagrangian. However, the major drawback of $L_2$ is that the action principle is not a well-posed one for arbitrary $f^i(q^A,\p q^A)$ (\textit{i.e.} one has to fix both $q^A(t)$ and $\p_i q^A(t)$ on the boundary, which spoils the well-posedness in the action principle). Thus, one can only apply the above HJ principle by adding suitable boundary term with $L_2$ to negate the problematic terms arising from the surface terms so that one has to fix only $q^A$ on the boundary. 

For a special case, where $f^i(q^A,\p q^A)$ is given by $f^i(q^A,\p q^A)=\sum_{A} q^A\p L_1/\p_i q^A=\sum_{A}q^AP^i_A$ (where, the summation ``$\sum_{A}$'' is over one/some/all components of generalized coordinates and the corresponding momenta), and, thereby, $L_2\longrightarrow L_P(q^A,\p q^A,\p^2 q^A)$ is defined as
\begin{align}
L_P(q^A,\p q^A,\p^2 q^A)=L_1(q^A,\p q^A)-\sum_A\p_i (q^AP^i_A),
\end{align}
  it can be shown that $L_1$ and $L_P$ yields the same equation of motion. However, for $L_1$ one has to fix $q^A~~ \forall A=\{1,2,...N\}$  while, for $L_P$, one has to fix (one/some/all) $P^i_A$ on the boundary. Thus, $L_1$ can be interpreted as the action of the coordinate space while the Lagrangian $L_P$, which is still not well-posed, can be interpreted as the action of the momentum space. Furthermore, for the Lagrangian $L_P$, the bulk part (which is the same as $L_1$ \textit{i.e.} $L_{bulk}=L_1$) and the surface part ($L_{sur}=-\sum_{A}\p_i (q^AP^i_A)$) are not independent. Instead, the bulk and the surface part are related by the following connection
\begin{align}
L_{sur}=-\sum_{A}\p_i \Big(q^A \frac{\p L_{bulk}}{\p_iq^A}  \Big)~. \label{HOLGEN}
\end{align}
Again, we emphasize that in the above relation \eqref{HOLGEN}, Einstein summation convention has been implied for the spacetime index $i$, whereas the summation $\sum_{A}$ is arbitrary, which can include one/some/all components of the generalized coordinates and the corresponding conjugate momenta.

The above general discussion is applicable for the action principle in general relativity (GR) and for the scalar-tensor theory as well. The Einstein-Hilbert Lagrangian $L_{EH}$ is given as $L_{EH}=R$, where $R$ is the Ricci-scalar. Note that $R$ contains both the first order as well as the second order derivative of the metric tensor $g_{ab}$. Also, the equation of motion obtained by the variation of the Einstein-Hilbert action (\textit{i.e.} the Einstein's equation) is also second order derivative. This is possible because, the Einstein-Hilbert Lagrangian $L_{EH}$ can be decomposed into the bulk part $L^{(EH)}_{bulk}$ and the surface part $L^{(EH)}_{sur}$, where $L^{(EH)}_{bulk}$ contains only the first order derivative of $g_{ab}$, whereas the surface term $L^{(EH)}_{sur}$, which is a total derivative term, contains the second order derivative of $g_{ab}$ (for more details, see \cite{gravitation}). Furthermore, it can be shown that the bulk part and the surface part are not independent. Instead, they are related by the ``Holographic relation'' \cite{Padmanabhan:2002jr,Padmanabhan:2002xm,Padmanabhan:2003gd,Padmanabhan:2004fq,Padmanabhan:2006fn,Mukhopadhyay:2006vu,Kolekar:2010dm,Bhattacharya:2022vta}, which is basically the analogous relation of eq \eqref{HOLGEN} . Thus, $L^{(EH)}_{bulk}$ corresponds to $L_1$ and the total Lagrangian $L_{EH}$ corresponds to $L_P$ of the above discussion. Moreover, both $L^{(EH)}_{bulk}$ and $L_{EH}$ yields the same equation of motion and $L^{(EH)}_{bulk}$ can be interpreted as the Lagrangian of the coordinate space, whereas the total Lagrangian $L_{EH}$ can be interpreted as the Lagrangian of the momentum space. 

In scalar-tensor theory, the situation is a bit different in the two different frames. In Einstein frame, which is very similar to the Einstein-Hilbert case, the total Lagrangian (as defined in eq. \eqref{SE}) can be decomposed into bulk and the surface term \textit{i.e.} $\sqrt{-\tilde{g}}\tilde{L}=\sqrt{-\tilde{g}} \t L_{bulk}+\t L_{sur}$, where 
\begin{align}
\tilde{L}_{bulk}=\frac{1}{16\pi}\tilde{g}^{ab}(\tilde{\Gamma}^{i}_{ja}\tilde{\Gamma}^{j}_{ib}-\tilde{\Gamma}^{i}_{ab}\tilde{\Gamma}^{j}_{ij}) -\frac{1}{2}\tilde{g}^{ab}\tilde{\nabla}_a\tilde{\phi}\tilde{\nabla}_b\tilde{\phi}-U(\tilde{\phi})~;
\label{QUAD}
\end{align}
and the surface term is given as
\begin{align}
\tilde{L}_{sur}=-\partial_c\tilde{P}^c~, 
\label{SUR}
\end{align}
where
\begin{align}
\tilde{P}^c=\frac{\sqrt{-\tilde{g}}}{16\pi}(\tilde{g}^{ck}\tilde{\Gamma}^i_{ki}-\tilde{g}^{ik}\tilde{\Gamma}^c_{ik})~. \label{PCTIL}
\end{align}
Furthermore, in Einstein frame, one can obtain the ``Holographic relation'' like the Einstein-Hilbert action, which is provided as follows.
\begin{align}
\tilde{L}_{sur}=-\partial_c\Big[\frac{\partial\sqrt{-\tilde{g}}\tilde{L}_{bulk}}{\partial\tilde{g}_{ij,c}}\tilde{g}_{ij}\Big]. \label{LEINTIL}
\end{align}
Thus, in Einstein frame, $\tmg \tilde{L}_{bulk}$ corresponds to $L_1$ and can be interpreted as the Lagrangian of the coordinate space. On the contrary, the total gravitational Lagrangian $\sqrt{-\tilde{g}}\tilde{L}$ can be interpreted as the Lagrangian of the momentum space and corresponds to $L_P$ of the above discussion. Both $\tmg \tilde{L}_{bulk}$ and $\sqrt{-\tilde{g}}\tilde{L}$ yields the same equation of motion. More importantly, $\tmg \tilde{L}_{bulk}$ provides a well-posed action principle (albeit not in a covariant way as the aforementioned decomposition in terms of bulk and surface part has been done in a non-covariant manner; for details see \cite{Bhattacharya:2017pqc,Bhattacharya:2020jgk,Bhattacharya:2022mnb}). But, the total Lagrangian $\sqrt{-\tilde{g}}\tilde{L}$ does not provide a well-posed action principle and one has to incorporate an additional boundary term in order to negate the contribution from the $\tilde{L}_{sur}$ on the boundary.

Let us now discuss the Jordan frame action in the light of the above discussions. The Jordan frame Lagrangian can be decomposed into the bulk and the surface terms \textit{i.e.} $\sqrt{-g}L=\sqrt{-g}L_{bulk}+L_{sur}$ where
\begin{eqnarray}
&&L_{bulk}=(1/16\pi)\Big[\Omega^2g^{ab}[\Gamma^i_{ja}\Gamma^j_{ib}-\Gamma^{i}_{ab}\Gamma^j_{ij}]-2\Omega^2g^{ab}\Gamma^i_{ab}(\partial_i\ln\Omega)
+2\Omega^2\Gamma^i_{ij}(\partial^j\ln\Omega)\Big]
\nonumber
\\
&&-\frac{4}{16\pi}\omega\Omega^2(\partial_i\ln\Omega)(\partial^i\ln\Omega)-\frac{V(\phi)}{16\pi\phi^2}~,
\label{bulk}
\end{eqnarray}
and
\begin{eqnarray}
&& L_{sur}=\frac{1}{16\pi}\partial_c[\Omega^2\sqrt{-g}(g^{ik}\Gamma^c_{ik}-g^{ck}\Gamma^m_{km})]~,
\label{3}
\end{eqnarray}
where the bulk part contains only the first order derivative of the field variables (\textit{i.e.} the metric tensor and the scalar field $\phi=\Omega^2$) and the surface part contains the second order derivative. Again, in this case, both the bulk Lagrangian $\sqrt{-g}L_{bulk}$ and the total Lagrangian $\sqrt{-g}L$ corresponds to the same equation of motion (for more details, see \cite{Bhattacharya:2017pqc,Bhattacharya:2020jgk,Bhattacharya:2022mnb}). Similar to the Einstein frame, in order to define a well-posed action principle for the total Lagrangian $\sqrt{-g}L$, one has to incorporate suitable boundary term. On the contrary, the bulk Lagrangian defines a well-posed action principle in a non-covariant manner. However, there exists a major difference as opposed to the Einstein frame. In this case, the earlier holographic relation does not hold \textit{i.e.}, the bulk part and the surface part are related to each other by the following relation \cite{Bhattacharya:2017pqc,Bhattacharya:2020jgk,Bhattacharya:2022mnb}
\begin{eqnarray}
&&L_{sur}=-\p_c\Big[\frac{\partial\sqrt{-g}L_{bulk}}{\partial g_{ab,c}}g_{ab}\Big]+\frac{3}{16\pi}\sqrt{-g}\square\phi~. \label{EQU}
\end{eqnarray}
It is the last term of the above equation \eqref{EQU} which spoils the holographic relation. As a result, the Jordan frame Lagrangian cannot be interpreted as the Lagrangian of the momentum space \textit{i.e.,} $\sqrt{-g}L$ corresponds to $L_2$ (not $L_P$) whereas the bulk part $L_{bulk}$ corresponds to $L_1$ of the above discussion. Therefore, there exists an inequivalence of the two frames even at the classical level. It has been investigated in the earlier works \cite{Bhattacharya:2018xlq} that this inequivalence at the action level translates to the major inequivalences of the two frames, especially at the thermodynamic level. However, the root of this inequivalence lies in eq. \eqref{ACTUAL}. Instead of discarding the $\square\phi$ term, if we incorporate it in the Jordan frame Lagrangian \textit{i.e.} we define the Jordan frame Lagrangian as $L'=L-3\square\phi/16\pi$ it can be shown that the action in the two frames are exactly equivalent \textit{i.e.} $\t{\mathcal{A}}=\mathcal{A}'$, where $\mathcal{A}'=\int_\nu d^4x\sqrt{-g}L'$. In addition, the holographic relation holds for the Lagrangian $L'$. The bulk and the surface part of $L'$ can be defined as $L'_{bulk}=L_{bulk}$ (\textit{i.e.} the bulk part of $L$ and $L'$ are the same) and 
\begin{align}
L'_{sur}=L_{sur}-\frac{3\mg}{16\pi}\square\phi=\frac{1}{16\pi}\partial_c\Big[\sqrt{-g}\Big\{\phi(g^{ik}\Gamma^c_{ik}-g^{ck}\Gamma^m_{km})-3g^{cd}\p_d\phi\Big\}\Big]~.
\end{align}
In this case, the bulk and the surface part are related to each other by the holographic relation \textit{i.e.}
\begin{align}
L'_{sur}=-\p_c\Big(\frac{\partial\sqrt{-g}L'_{bulk}}{\partial g_{ab,c}}g_{ab}\Big)~. \label{HOLNEW}
\end{align}
Thus, the inequivalence can be removed and $L'$ can be interpreted as the Lagrangian of the momentum space (\textit{i.e.} $L_P$ of the above discussion). In our subsequent analysis, when we obtain the boundary term and, thereby, establish the well-posedness, we find that the same argument will be valid. The gravitational action of the Einstein frame can be interpreted as the action of momentum space, whereas the action of the Jordan frame cannot be interpreted as the same unless the $\square\phi$ term is accounted.

In order to obtain a well-posed action principle we, therefore, have two choices: We obtain the equation of motion solely from the bulk Lagrangians $\tmg \tilde{L}_{bulk}$ and $\sqrt{-g}L_{bulk}$ respectively. This method is non-covariant and has been explored earlier \cite{Bhattacharya:2017pqc,Bhattacharya:2020jgk,Bhattacharya:2022mnb} and we have briefly discussed above. Secondly, we can incorporate suitable boundary terms along with the gravitational Lagrangians in order to obtain a well-posed action principle (similar to the Gibbons-Hawking-York boundary term which is added along with the Einstein-Hilbert Lagrangian). In the following, we shall explore second route in detail. This method is covariant, yet foliation-dependent. In addition, this method helps us to obtain the surface energy-momentum and to define the quasi-local charges using the Brown-York formalism. As we have discussed, the Brown-York formalism and obtaining the suitable boundary term for a well-posed action principle is foliation-dependent. Therefore, in the following section, we analyze for the timelike (or spacelike) surfaces. Thereafter, the null-surface will be treated separately, which is more non-trivial.

Before proceeding to the next section, let us summarise the discussion provided in this section as follows:
 Firstly, both the gravitational actions in the two frames ($\mathcal{A}$ and $\t{\mathcal{A}}$) are not well-posed. In order to obtain the equation of motion, there are two possibilities: (i) One can follow a non-covariant approach whereby one decomposes the gravitational action into bulk and surface part where one obtains the equation of motion from only the bulk part of the action. (ii) One can add a suitable boundary term with the gravitational action, which cancels the problematic terms on the boundary. This method is covariant yet foliation-dependent and has been followed in the subsequent discussions of the paper.  
 Secondly, the usual actions in the two frames ($\mathcal{A}$ and $\t{\mathcal{A}}$) are not exactly equivalent. Not only they are mathematically inequivalent, they carry different interpretations. While $\t{\mathcal{A}}$ can be interpreted as the action of the momentum space, we cannot draw the same conclusion for $\mathcal{A}$. In addition, the holographic relation cannot be obtained in the Jordan frame, whereas the same is available for the Einstein frame.
Thirdly, the aforementioned inequivalence can be removed by redefining the action of the Jordan frame as $\mathcal{A}'$, whereby one can obtain the holographic relation and can interpret $\mathcal{A}'$ as the action of the momentum space. Furthermore, the earlier analysis from the viewpoint of black hole thermodynamics also lauds $\mathcal{A}'$.


\section{Boundary term and the Brown-York formalism on a timelike hypersurface} \label{sectimelike}
 In order to define a well-posed action principle, it was found that one can add several boundary terms along with the gravitational action \cite{Charap:1982kn}, which was originally found in GR.  Recently, it was argued \cite{Padmanabhan:2014lwa,Parattu:2015gga,Parattu:2016trq,Chakraborty:2017zep,Chakraborty:2018dvi} that we should not pre-impose the boundary term, in order to define a well-posed action principle. Instead, the action principle itself should tell us what boundary term one should add so that if the action is well-posed, the number of degrees of freedom which are required to be fixed on the boundary, will correspond to the number of true degrees of freedom in the theory (for details, see \cite{Parattu:2015gga}). Although the gravitational actions (in both the frames) of scalar-tensor theory are also not well-posed (as we have discussed in the earlier section), a discussion on such spirit (\textit{i.e.} letting the action principle decide what to be fixed on the boundary) is missing in literature. In addition, the boundary term for a null surface has not been defined earlier for ST gravity. In our following discussion, we have completed this picture. Furthermore, we obtain the quasi-local Brown-York parameters in both the frames and we compare how they are related in both the frames.

In this section, we shall discuss the issues related to the variation of the action, boundary terms and the Brown-York formalism for timelike/ spacelike hypersurface. In the following section, the null surface is treated separately due to the fact that the usual formalism of timelike (or spacelike) surface does not work for the null surface. For simplicity, we first perform the analysis in the Einstein frame, which is exactly similar to the Einstein's gravity (GR) and then the discussions on the Jordan frame will follow.

\subsection{Einstein frame}
In Einstein frame, the non-minimal coupling is not present and, therefore, the analysis is simpler. In this case, the variation of the gravitational action is given as
\begin{eqnarray}
&&\d\tilde{\mathcal{A}}=\int_\nu \delta(\sqrt{-\tilde{g}}\tilde{L})d^4x=\int_\nu
\sqrt{-\t{g}}\Big[\t{E}_{ab}\delta{\t{g}^{ab}}+\t{E}_{(\t{\phi})}\delta{\t{\phi}}
+\t{\na}_a\t{\T}^a (\t{q},\delta{\t{q}})\Big]d^4x~, \label{VAR2}
\end{eqnarray}
where $\t{q}\in \{\t{g}_{ab}, \t{\phi}\}$. The exact expressions of  $\t{E}_{ab}$, $\t{E}_{(\t{\phi})}$ and $\t{\T}^a (\t{q},\delta{\t{q}})$ are given as follows
\begin{align}
 \t{E}_{ab}=\frac{\tilde{G}_{ab}}{16\pi}-\frac{1}{2}\tilde{\nabla}_a\tilde{\phi}\tilde{\nabla}_b\tilde{\phi}+\frac{1}{4}\tilde{g}_{ab}\tilde{\nabla}^i\tilde{\phi}\tilde{\nabla}_i\tilde{\phi}+\frac{1}{2}\tilde{g}_{ab}U(\tilde{\phi})~;
\no 
\\
\t{E}_{(\t{\phi})}=\t{\na}_a\t{\nabla}^a\tilde{\phi}-\frac{dU}{d\tilde{\phi}}~;\ \ \ \ \ \ \ \ \ \ \ \
\no 
\\
\textrm{and} \ \ \ \ \ \ \ \ \ \ \ \ \ \ \ \ \ \ \ \ \ \ \ \ \
\no 
\\
\t{\T}^a (\t{q},\delta{\t{q}})=\frac{\delta\tilde{v}^a}{16\pi}-(\tilde{\nabla}^a\tilde{\phi})\delta\tilde{\phi}~. \ \ \ \ \ \  \label{EXACTEXEIN}
\end{align}
Here, $\d\t v^a=2\t P^{ibad}\t\na_b\d \t g_{id}$, with $\t P^{abcd}=(\t g^{ac}\t g^{bd}-\t g^{ad}\t g^{bc})/2$. One can identify that $\t{E}_{ab}=0$ and $\t{E}_{(\t{\phi})}=0$ corresponds to the equations of motion of the fields $\t g^{ab}$ and $\t\phi$. However, one has to properly deal with the boundary term in order to obtain a well-posed action principle. Dealing with the boundary term in the present case is problematic (as is the case for the Einstein-Hilbert action). The last term of $\t{\T}^a (\t{q},\delta{\t{q}})$ (\textit{i.e.} the term containing $\d\t\phi$) vanishes if we fix $\t\phi$ on the boundary. On the contrary, $\d\t v^a$ vanishes only if we fix both the metric tensor ($\t g_{ab}$) as well as its first order derivative ($\t\p_i\t g_{ab}$) on the boundary, which is not physical. To get rid of such situation one can add a boundary term such a way that one has to fix $\t g_{ab}$ or $\t\p_i\t g_{ab}$ on the boundary. We do not pre-impose the boundary term with the gravitational Lagrangian. Instead we let the action principle decide what boundary term should be added with the gravitational action so that the minimal information are required to be fixed on the boundary \footnote{As per the argument provided in \cite{Padmanabhan:2014lwa,Parattu:2015gga,Parattu:2016trq,Chakraborty:2017zep,Chakraborty:2018dvi}, the number of quantities which are required to be fixed should correspond to the true degrees of freedom in the theory. In GR, it has been found that six components of the induced metric are required to be fixed on the boundary. In the present case, we see that, the scalar field is required to be fixed on the boundary in addition to the six components of the metric tensor. Thus, in ST gravity, we have additional scalar degrees of freedom which acts as the true degrees of freedom in the theory (as implied in the name ``scalar tensor theory'').}. This method is in contrast to the usual approach where one adds a boundary term and shows that the problematic terms go away.

We first consider a three-dimensional surface (say $\psi=$ const.), upon which $\t r_a$ is unit normal \textit{i.e.} $\t r^a\t r_a=\epsilon$ (where $\epsilon=-1$ for timelike normal or $\epsilon=+1$ for spacelike normal on the different parts of the boundary $\p\nu$. In order to keep the generality of the surface, we keep it $\epsilon$). Our aim is to obtain the boundary term which makes the action principle well-posed on this surface. One can construct the following induced metric on this surface
\begin{align}
\t h_{ab}^{(\t r)}=\t g_{ab}-\epsilon \t r_a\t r_b~,
\end{align}
 which acts as the projection tensor, that projects everything on to the tangent plane of the surface (as $\t h_{ab}^{(\t r)}\t r^a=0$ and $\t h^{(\t r)a}_b\t h^{(\t r)b}_c=\t h^{(\t r)a}_c$). On this surface, the contribution from the boundary term (\textit{i.e.} the last term of Eq. \eqref{VAR2}) is given as 
\begin{align}
\t{\mathcal{B}}=\int_{\p\nu}\epsilon\sqrt{\t h^{(r)}}\t r_a\t\T^a d^3x=\int_{\p\nu}\epsilon\sqrt{\t h^{(r)}}\Big[\frac{1}{16\pi}\t r_a\d\t v^a-(\t r_a\t\na^a\t\phi)\d\t\phi\Big]d^3x~. \label{bountil}
\end{align}
The above expression in \eqref{bountil} is obtained using Stoke's theorem \textit{i.e.} by changing the volume integration to surface integration. Our goal is to identify the surface term which is required to be subtracted from $\t{\mathcal{B}}$ in order to obtain a well-posed action principle. Schematically, we expect the following expression of $\t{\mathcal{B}}$
\begin{align}
\t{\mathcal{B}}=\int_{\p\nu}d^3x~\Big[ \d\textrm{(Boundary Term)}+\textrm{(Conjugate Momentum)}\d\textrm{(Variables to be fixed)}
\no 
\\
+\textrm{Total Derivative Term}\Big]~. \label{schmatic}
\end{align}
To obtain such expression we use the following geometrical identity (\cite{Padmanabhan:2014lwa}, also see the appendix \ref{proofidentity})
\begin{align}
X_a\d v^a=\na_a\Big(\d X^a+g^{ab}\d X_b\Big)-\d(2\na_a X^a)+\na_a X_b\d g^{ab}~, \label{identity}
\end{align}
where $X_a$ is any vector (timelike/ spacelike/null) and $\delta v^a=2P^{ibad}\na_b\d g_{id}$. Using the above identity \eqref{identity}, the boundary contribution (as defined in \eqref{bountil}) finally yields (the mathematical details are the same as of GR \cite{Padmanabhan:2014lwa})
\begin{align}
\t{\mathcal{B}}=\int_{\p\nu}\epsilon\Bigg[\frac{1}{16\pi}\Bigg(\sqrt{\t h^{(r)}}~^{(\t r)}\t D_a\Big(\t h^{(r)a}_i\t r_j\d\t g^{ij}\Big)+2\d\Big(\sqrt{\t h^{(r)}}\t\theta^{(\t r)}\Big)
\nonumber
\\
+\sqrt{\t h^{(r)}}\Big(\t\theta^{(\t r)}\t h_{ab}^{(\t r)}-\t\theta^{(\t r)}_{ab}\Big)\d\t h^{ab}_{(r)}\Bigg)-\sqrt{\t h^{(r)}}(\t r_a\t\na^a\t\phi)\d\t\phi\Bigg]d^3x~,\label{bounrtil}
\end{align}
where the extrinsic curvature is defined as
\begin{align}
\t\theta^{(\t r)}_{ab}=-\t h_a^{(\t r)i}h_b^{(\t r)j}\t\na_i\t r_j=-\t h_a^{(\t r)i}\t\na_i\t r_b
\no 
\\
=-\t\na_a\t r_b+\epsilon \t r_a \t a^{(\t r)}_b~, \label{extrintil}
\end{align}
and the covariant derivative operator on $\p \nu$, which is compatible with $\t h_{ab}^{(\t r)}$ is defined as
\begin{align}
^{(\t r)}\t D_a\t A_b=\t h_a^{(\t r)i}h_b^{(\t r)j}\t\na_i\t A_j~. \label{threedertil}
\end{align}

 In eq. \eqref{bounrtil}, we finally obtain the desired expression of the form \eqref{schmatic}. One can identify the boundary term which is required to be added to the gravitational action $\t{\mathcal{A}}$ in order to obtain a well posed action principle on the surface $\p\nu$, which is given as
\begin{align}
\t{\mathcal{A}}_{sur}=-\frac{\epsilon}{8\pi}\int_{\p\nu}\sqrt{\t h^{(r)}}\t\theta^{(\t r)}d^3x\equiv -\frac{1}{8\pi}\int_{\nu}\tmg\tna_a\Big(\t\theta^{(\t r)}\t r^a\Big)d^4x~,\label{surrtil}
\end{align}
where the last expression is the equivalent bulk term corresponding to the surface term. The first term of \eqref{bounrtil} is a total derivative term on the three surface, which can be neglected. In scalar-tensor gravity, the dynamical parameters are the metric tensor $\t g_{ab}$ and the scalar field $\t\phi$; which consist of eleven independent components in total (ten are the independent components of the metric tensor and the scalar field). As we can see, all the components of the metric tensor (\textit{i.e.} ten independent components) are not required to be fixed. Instead, we require to fix only six independent components of $\t h^{ab}_{(\t r)}$ ($\t h^{ab}_{(\t r)}$ has six independent components as it is a symmetric tensor with the constraint $ \t r_ah^{ab}_{(\t r)}=0$). Thus, in the present case, we require to fix seven components on the boundary (\textit{i.e.} six independent components of $\t h^{ab}_{(\t r)}$ and the scalar field $\t\phi$) whereas, in GR, we had to fix only six independent components of the induced metric.

 Thus, for this generic timelike/spacelike surface, the well-posed action will be $\t{\mathcal{A}}_{WP}=\t{\mathcal{A}}+\t{\mathcal{A}}_{sur}$ and the variation of $\t{\mathcal{A}}_{WP}$ (where ``WP'' stands for ``well-posed'') will be given as
\begin{align}
\d\t{\mathcal{A}}_{WP}=\int_\nu
\sqrt{-\t{g}}\Big[\t{E}_{ab}\delta{\t{g}^{ab}}+\t{E}_{(\t{\phi})}\delta{\t{\phi}}\Big]d^4x+\epsilon\int_{\p\nu}\sqrt{\t h^{(r)}}\Bigg[~^{(\t r)}\t D_a\t T_{(\t r)}^a+\t\Pi^{(r)}_{ab}\d\t h^{ab}_{(r)}-(\t r_a\t\na^a\t\phi)\d\t\phi\Bigg]d^3x~, \label{AWPTIL}
\end{align}
where $\t T_{(\t r)}^a=\t h^{(r)a}_i\t r_j\d\t g^{ij}/16\pi$, and
\begin{align}
\t\Pi^{(r)}_{ab}=\frac{1}{16\pi}\Big(\t\theta^{(\t r)}\t h_{ab}^{(\t r)}-\t\theta^{(\t r)}_{ab}\Big)~, \label{CONOEIN}
\end{align}
which is the canonical momentum conjugate to $\t h^{ab}_{(r)}$~. From eq. \eqref{AWPTIL}, it can be concluded that $\t{\mathcal{A}}_{WP}$ can be interpreted as the action of the coordinate space. One can now ask whether $\t{\mathcal{A}}$ can be interpreted as the action of the momentum space as discussed in section \ref{sect2}.  With the above definition of canonical momentum in eq. \eqref{CONOEIN}, the surface term $\t{\mathcal{A}}_{sur}$ can be defined as 
\begin{align}
\t{\mathcal{A}}_{sur}=-\epsilon\int_{\p\nu}\sqrt{\t h^{(r)}}\t\Pi_{(r)}^{ab}\t h^{(r)}_{ab}d^3x~.
\end{align}
Therefore, another expression of the variation of the gravitational action $\t{\mathcal{A}}=\t{\mathcal{A}}_{WP}-\t{\mathcal{A}}_{sur}$ is given as
\begin{align}
\d \t{\mathcal{A}}=\int_\nu
\sqrt{-\t{g}}\Big[\t{E}_{ab}\delta{\t{g}^{ab}}+\t{E}_{(\t{\phi})}\delta{\t{\phi}}\Big]d^4x+\int_{\p\nu}\epsilon\Bigg[\sqrt{\t h^{(r)}}~^{(\t r)}\t D_a\t T_{(\t r)}^a+\t h^{ab}_{(r)}\d\Big(\sqrt{\t h^{(r)}}\t\Pi^{(r)}_{ab}\Big)
\no 
\\
-\sqrt{\t h^{(r)}}(\t r_a\t\na^a\t\phi)\d\t\phi\Bigg]d^3x~.
\end{align}
Thus, we again find that the gravitational action $\t{\mathcal{A}}$ can be interpreted as the action of the momentum space, where the conjugate momenta (rather $\sqrt{\t h^{(r)}}\t\Pi^{(r)}_{ab}$) are required to be fixed on the boundary. This agrees with the discussions we provided in section \ref{sect2}, where we had mentioned that the whole gravitational action can be interpreted as the action of the momentum space; whereas its bulk decomposition (which is a non-covariant term) can be interpreted as the action of the coordinate space. In this section, we find that the above well-posed action $\t{\mathcal{A}}_{WP}$, which is covariant yet foliation-dependent, can again be interpreted as the action of the coordinate space. Thus, analysis from different directions converge towards the similar conclusion.

The above analysis is performed for a particular surface, which is defined by the normal $\t r$. We consider that the whole four-dimensional manifold ($\mathcal{M}$) has the boundary $\p\mathcal{M}$, which consists of initial and final spacelike hypersurfaces (defined by $t_i=$const. and $t_f=$const. respectively) and a three-dimensional timelike boundary $^3\mathcal{B}$.  We further assume that the normal on $^3\mathcal{B}$ is denoted as $\t s_a$ ($\t s_a\t s^a=+1$, \textit{i.e.} $\epsilon=+1$ for $^3\mathcal{B}$) and the normal to a generic $t=$const. hypersurface (say $\Sigma$) is denoted as $\t n_a$ ($\t n_a\t n^a=-1$, \textit{i.e.} $\epsilon=-1$ for $\Sigma$). For simplicity, we consider the hypersurface foliations $\Sigma$ and $^3\mathcal{B}$ are orthogonal to each other (\textit{i.e.} $\t n^a \t s_a=0$). Since our final goal is to obtain the boundary term and to define the Brown-York quasi-local parameters, this does not break any generality. The boundary of $\Sigma$ will be a two-surface which we denote as $\mathcal{B}$, where $\mathcal{B}=~^3\mathcal{B}\cap \Sigma$. On these surfaces, the induced metric and the extrinsic curvature are provided in the following table.

\begin{table}[H]
\begin{tabular}{ |p{3 cm}|p{3 cm}|p{3 cm}|p{3 cm}|p{3 cm}| } 
 \hline
 \textbf{Surface} & \textbf{Normal(s)} & \textbf{Induced metric} & \textbf{Extrinsic curvature} & \textbf{Trace of the extrinsic curvature} \\ 
 \hline
 ~~~~~~~~~~~$\Sigma$ & ~~~~~~~~~~~$\t n_a$ & $\t h_{ab}=\t g_{ab}+\t n_a\t n_b$ & ~$\t K_{ab}=-\t h^i_a\t \na_i\t n_b$& $\t K=-\t \na_a \t n^a$\\
 \hline 
 ~~~~~~~~~~$^3\mathcal{B}$ & ~~~~~~~~~~~$\t s_a$ & $\t \gamma_{ab}=\t g_{ab}-\t s_a\t s_b$ &~$\t \K_{ab}=-\t \gamma^i_a\t \na_i\t s_b$ & $\t \K=-\t \na_a\t s^a$ \\ 
 \hline
 ~~~~~~~~~~~$\mathcal{B}$ & ~~~~~~$\t n_a$ and $\t s_a$ &$\t q_{ab}= \t g_{ab}+ \t n_a \t n_b- \t s_a \t s_b$ &~$\t k_{ab}=- \t q^i_a \t q^j_b\t \na_i \t s_j$ &$ \t k=- q^{ij}\t \na_i \t s_j=(\t \K_{ab}-\t \K \t \gamma_{ab}) \t n^a \t n^b$\\
 \hline

\end{tabular}
 \caption{\label{EINTABLE} Normals and extrinsic curvatures of different surfaces of the manifold in the Einstein frame}
 \end{table}

With all the above definitions (provided in Table \ref{EINTABLE}), we can now define the well-posed action for the manifold $\mathcal{M}$, which is the gravitational action added with the boundary terms for each three-surfaces  \textit{i.e.,}
\begin{align}
\tilde{\mathcal{A}}_{tot}=\tilde{\mathcal{A}}-\frac{1}{8\pi}\int_{^3\mathcal{B}}\sqrt{\t\gamma}\tK d^3x+\frac{1}{8\pi}\int_{t_i}^{t_f}\sqrt{\t h}\t Kd^3x~,
\end{align}
where $\int_{t_i}^{t_f}$ is the shorthand of $\int_{t_f}-\int_{t_i}$. The variation of the above action yields
\begin{align}
\d\tilde{\mathcal{A}}_{tot}=\int_{\mathcal{M}}
\sqrt{-\t{g}}\Big[\t{E}_{ab}\delta{\t{g}^{ab}}+\t{E}_{(\t{\phi})}\delta{\t{\phi}}
\Big]d^4x+\int_{^3\mathcal{B}}\sqrt{\t\gamma}\Big(\t{\mathcal{D}}_a\t{\mathcal{T}}^a+\t\Pi_{ab}\d\t\gamma^{ab}-(\t s_a\t\na^a\t\phi)\d\t\phi\Big)d^3x
\nonumber
\\
-\int_{t_i}^{t_f}\sqrt{\t h}\Big(\t{D}_a\t{T}^a+\t P_{ab}\d\t h^{ab}-(\t n_a\t\na^a\t\phi)\d\t\phi\Big)d^3x~, \label{varBYtil}
\end{align}
where $\t{\mathcal{D}}_a$ and $\t{D}_a$ are the covariant derivative operators compatible with $\t\gamma_{ab}$ and $\t h_{ab}$ respectively and 
\begin{eqnarray}
&&\t\Pi_{ab}=\frac{1}{16\pi}\Big[\tK\t\gamma_{ab}-\tK_{ab}\Big]~~~~~~~~~ \t P_{ab}=\frac{1}{16\pi}\Big[\t K\t h_{ab}-\t K_{ab}\Big]
\nonumber
\\
&&\t{\mathcal{T}}^a=\frac{1}{16\pi}\t\gamma_i^a\t s_j\d\t g^{ij}~~~~~~~~~~~~~~~~~\t T^a=\frac{1}{16\pi}\t h_i^a\t n_j \d\t g^{ij}      ~.
\end{eqnarray}
The total-derivative terms $\t{\mathcal{D}}_a\t{\mathcal{T}}^a$ and $\t{D}_a\t{T}^a$ can be ignored. Thus, we found that on each surface, one has to fix the corresponding induced metric and the scalar field $\t\phi$. This allow us to define the boundary stress-tensor (or the Brown-York tensor), which is defined as \cite{Brown:1992br,Brown:2000dz}
\begin{align}
\t T_{ab}^{(BY)}=-\frac{2}{\sqrt{\t\gamma}}\frac{\d\tilde{\mathcal{A}}_{tot}}{\d\t\gamma^{ab}}=-2\t\Pi_{ab}=\frac{1}{8\pi}\Big[\tK_{ab}-\tK\t\gamma_{ab}\Big] \label{BYEMTIL}
\end{align} 
From this surface tensor, one can obtain the quasi-local parameters of the surface. The quasilocal surface energy density (also known as the Brown-York quasilocal energy density, $\t \epsilon^{(BY)}$), surface tangential momentum density ($\t j^a$) and the spatial stress ($\t s^{ab}$) are defined as 
\begin{align}
\t \epsilon^{(BY)}=\t T_{ab}^{(BY)} \t n^a\t n^b=\frac{\t k}{8\pi}~,
\no 
\\
\t j^a=\t T_{bc}^{(BY)} \t n^b \t q^{ac}~,
\no 
\\
\t s^{ab}=\t q^{ac} \t q^{bd}\t T_{cd}^{(BY)}~.
\end{align}
Furthermore, the Brown-York energy is defined as the energy density integrated over the two surface \textit{i.e.},
\begin{align}
\t E^{(BY)}=\int_{\mathcal{B}}\sqrt{\t q}\Big(\t \epsilon^{(BY)}-\t \epsilon^{(BY)}_0\Big)d^2x
\end{align}
where $\t\epsilon^{(BY)}_0$ corresponds to the contribution from the reference frame \cite{Brown:1992br,Brown:2000dz} and $\t q$ is the determinant of the induced metric $\t q_{ab}$~. When there is a rotational Killing vector $\t\xi^a$, the corresponding angular momentum is obtained from the BY tensor as 
\begin{align}
\t J=\int_{\mathcal{B}}\sqrt{\t q}~\t j_a\t\xi^a~d^2x.
\end{align}
 Furthermore, the spatial stress $\t s^{ab}$ can be expressed as the stress tensor of a viscous fluid of the following form 
\begin{align}
\t s^{ab}=\Big[2\t\eta\t\sigma^{ab}+\t q^{ab}(\t\zeta\t \Theta-\t P)\Big]~, \label{fluidtil}
\end{align}
where the shear tensor $\t\sigma^{ab}$ is the traceless part of $\t k^{ab}$, \textit{i.e.} $\t \sigma^{ab}=\t k^{ab}-\t k\t q^{ab}/2~.$ On the other hand, the bulk viscosity ($\t\Theta$) is identified as $\t\Theta=\t k$, and the pressure term is identified as $\t P=-\t a^{(\t n)}_i\t s^i/8\pi$ where $\t a^{(\t n)}_i=\t n^a\t \na_a\t n_i$~. In addition, the shear viscosity coefficient $\t\eta$ is obtained as $\eta=1/16\pi$, and the bulk viscosity coefficient is obtained as $\t\zeta=-1/16\pi$. The above expressions interprets the two surface $\mathcal{B}$ as the membrane of a two-dimensional viscous fluid, which is the central theme of the ``membrane paradigm'' \cite{Price:1986yy,Parikh:1997ma}. One striking aspect in this case is the negative value of the bulk viscosity coefficient, which indicates the instability against a perturbation which triggers expansion or contraction.  Furthermore, one can show that the ratio of the shear viscosity $\t\eta$ to the entropy density $\t s=1/4$ saturates the Kovtun-Son-Starinets (KSS) bound \textit{i.e.}
\begin{align}
\frac{\t\eta}{\t s}=\frac{1}{4\pi}~.
\end{align} 

We now move on to the analysis in the original frame \textit{i.e.} the Jordan frame. Thereafter, we can compare the two frames. Due to the presence of the non-minimal coupling, the analysis in the Jordan frame is more non-trivial, which is provided as follows.

\subsection{Jordan frame}
Earlier, we have shown how the well-posed action can be formulated in the Einstein frame. In that case, we have shown that the proper boundary term can be obtained from the action principle itself. In the following, we follow the same principle for the gravitational action in the Jordan frame. The variation of the gravitational action \eqref{SJ} is given as

\begin{align}
\d\mathcal{A}=\int_\nu\delta(\sqrt{-g}L)d^4x=\int_\nu\sqrt{-g}\Big(E_{ab}\delta g^{ab}+E_{(\phi)}\d\phi+\na_a\T^a(q,\d q)\Big)d^4x~, \label{DELL'}
\end{align}
where $q\in \{g_{ab}, \phi\}$ and the exact expressions of $E_{ab}$, $E_{(\phi)}$, and $\T^a (q,\delta{q})$ are provided as follows
\begin{eqnarray}
&& E_{ab}=\frac{1}{16\pi}\Big[\phi G_{ab}+\frac{\omega}{2\phi}\nabla_i\phi\nabla^i\phi g_{ab}-\frac{\omega}{\phi}\nabla_a\phi\nabla_b\phi
+\frac{V}{2}g_{ab}-\nabla_a\nabla_b\phi+\nabla_i\nabla^i\phi g_{ab}\Big]~;
\no
\\
&& E_{(\phi)}=\frac{1}{16\pi}\Big[R+\frac{1}{\phi}\frac{d\omega}{d\phi}\nabla_i\phi\nabla^i\phi +\frac{2\o}{\phi}\square\phi-\frac{dV}{d\phi}-\frac{\omega}{\phi^2}\nabla_a\phi \nabla^a\phi\Big]~;
\no 
\\
&&\ \ \ \ \ \ \ \ \ \ \ \ \ \ \ \ \ \ \ \ \ \ \ \ \ \ \ \ \ \ \ \ \ \ \ \ \ \ \ \textrm{and} 
\no 
\\
&&\T^a (q,\delta{q})=\frac{1}{16\pi}\Big[-2g^{ab}\frac{\omega}{\phi}(\nabla_b\phi) \delta\phi +\phi \delta v^a-2(\nabla_b\phi)p^{iabd}\delta g_{id}\Big]~, \label{THETA}
\end{eqnarray}
where $\d v^a=2P^{ibad}\na_b g_{id} $, with $P^{abcd}=(g^{ac}g^{bd}-g^{ad}g^{bc})/2$. Again, $E_{ab}$ and $E_{(\phi)}$ corresponds to the dynamical equation of the fields $g_{ab}$ and $\phi$ respectively. However, the action $\mathcal{A}$ is not a well-posed one for the arguments provided earlier. As we did for the Einstein frame, we first obtain the surface term for a generic surface which makes the action principle a well-posed one. We consider a three-dimensional generic hypersurface $\psi=$ const., upon which $r_a$ is the unit normal \textit{i.e.} $r^a r_a=\epsilon=\pm 1$ depending on timelike or spacelike hypersurface. It can be shown that $\t r_a$ and $r_a$ are connected as (remember $\t r_a$ is the normal in the Einstein frame on the same surface)
\begin{align}
\t r_a=\sqrt{\phi} r_a ~~~~~~~~~~~~~\textrm{and}~~~~~~~~~~~~~~~~~ \t r^a=\frac{1}{\sqrt{\phi}} r^a~.
\end{align}
On this surface (with normal $r_a$), the projection tensor can be defined as
\begin{align}
h_{ab}^{(r)}=g_{ab}-\epsilon r_a r_b~.
\end{align}

Again, on this surface, we obtain the contribution from the boundary term, which is provided as
\begin{align}
\mathcal{B}=\int_{\p\nu}\epsilon\sqrt{h^{(r)}}r_a\T^a d^3x=\frac{1}{16\pi}\int_{\p\nu}\epsilon\sqrt{h^{(r)}}\Big[\phi r_a\delta v^a-2r_a(\nabla_b\phi)p^{iabd}\delta g_{id}-\frac{2\omega}{\phi} (r^i\na_i\phi)\d\phi\Big]d^3x~, \label{btotjor}
\end{align}
where the volume integration has been reduced to the surface integration using Stoke's theorem. Again, our goal is to re-structure the above expression of $\mathcal{B}$ in terms of the structure as provided in eq. \eqref{schmatic}. We provide the final expression (detail calculations can be found in appendix \ref{appen2}), which is given as
\begin{align}
\mathcal{B}=\frac{\epsilon}{16\pi}\int_{\p\nu}\Bigg[\sqrt{h^{(r)}}~^{(r)}D_a\Big(\phi h^{(r)a}_i r_j\d g^{ij}\Big)+2\delta\Big(\sqrt{h^{(r)}}\phi \theta^{(r)}\Big)-2\sqrt{h^{(r)}}\Big(\frac{\omega}{\phi}(r^i\na_i\phi)+\theta^{(r)}\Big)\d\phi
\no 
\\
+\sqrt{h^{(r)}}\Big(\phi\Big\{-\theta^{(r)}_{ab} + \theta^{(r)}h^{(r)}_{ab}\Big\}-r^i\na_i\phi h^{(r)}_{ab}\Big)\d h^{(r)ab}\Bigg]d^3x~. \label{boufin}
\end{align}
 The first term, being a total derivative term on the three surface can be ignored. Also, from the above expression \eqref{boufin}, one can determine the boundary term which is required to be added to the gravitational action in order to define a well-posed action principle. The boundary term can be identified as
\begin{align}
\mathcal{A}_{sur}=-\frac{\epsilon}{8\pi}\int_{\p\nu}\sqrt{h^{(r)}}\phi \theta^{(r)}d^3x\equiv -\frac{1}{8\pi}\int_{\nu}\mg\na_a\Big(\phi \theta^{(r)}r^a\Big)d^4x~, \label{ABJOR}
\end{align}
 where the last expression is the equivalent bulk term corresponding to the surface term. More importantly, we find that if we add the surface term $\mathcal{A}_{sur}$ with the gravitational action $\mathcal{A}$, we obtain a well-posed action principle. In addition, we find that we do not need to fix all the ten independent components of the metric tensor $g_{ab}$ on the boundary. Instead, we need to fix $h^{(r)ab}$ which has six independent components (as $h^{(r)ab}$ is a symmetric tensor with the condition $h^{(r)ab}r_a=0$). Also, we require to fix $\phi$ on the boundary. Thus, like the Einstein frame, we require to fix seven independent components on the boundary. Moreover, the variation of the well-posed action (\textit{i.e.} $\mathcal{A}_{WP}=\mathcal{A}+\mathcal{A}_{sur}$) will be given as
 \begin{align}
\d\mathcal{A}_{WP}=\int_\nu\sqrt{-g}\Big(E_{ab}\delta g^{ab}+E_{(\phi)}\d\phi\Big)d^4x+\epsilon\int_{\p\nu}\sqrt{h^{(r)}}\Bigg[~^{(r)}D_a T^a_{(r)}+\Pi_{ab}^{(r)}\d h^{ab}_{(r)}+\Pi_{(\phi)}^{(r)}\d\phi\Bigg]d^3x \label{awpccord}
 \end{align}
 where $T^a_{(r)}=\phi h^{(r)a}_i r_j\d g^{ij}/16\pi$. From the above equation \eqref{awpccord}, one can identify that the well-posed action $\mathcal{A}_{WP}$ can be interpreted as the action of the coordinate space and the conjugate momenta are defined as
\begin{align}
\Pi_{ab}^{(r)}=\frac{1}{16\pi}\Big[\phi\Big(\theta^{(r)}h^{(r)}_{ab}-\theta^{(r)}_{ab}\Big)-r^i\na_i\phi h^{(r)}_{ab}\Big] ~~~~~~~\Pi_{(\phi)}^{(r)}=-\frac{1}{8\pi}\Big[\theta^{(r)}+\frac{\omega}{\phi}(r^i\na_i\phi)\Big]~.
\end{align}
In order to check whether the gravitational action $\mathcal{A}$ can be interpreted as the action of the momentum space, we obtain the surface term in terms of the conjugate momentum $\Pi_{ab}^{(r)}$, which is given as 
\begin{align}
\mathcal{A}_{sur}=-\epsilon\int_{\p\nu}\sqrt{h^{(r)}}\Big[\Pi^{ab}_{(r)}h_{ab}^{(r)}+\frac{3}{16\pi}r^i\na_i\phi\Big]d^3x~.
\end{align}
Unlike Einstein frame, here the surface term is not obtained only in terms of the product of $\Pi^{ab}_{(r)}$ and $h_{ab}^{(r)}$.
Therefore, although the well-posed action $\mathcal{A}_{WP}$ can be interpreted as the action of the coordinate space, the gravitational action $\mathcal{A}(=\mathcal{A}_{WP}-\mathcal{A}_{sur})$ cannot be interpreted as the action of the momentum space as
\begin{align}
\d\mathcal{A}=\int_\nu\sqrt{-g}\Big(E_{ab}\delta g^{ab}+E_{(\phi)}\d\phi\Big)d^4x+\epsilon\int_{\p\nu}\Bigg[\sqrt{h^{(r)}}\Big(~^{(r)}D_a T^a_{(r)}+\Pi_{(\phi)}^{(r)}\d\phi\Big)+h_{ab}^{(r)}\d\Big(\sqrt{h^{(r)}}\Pi^{ab}_{(r)}\Big)
\no 
\\
+\frac{3}{16\pi}\d\Big(\sqrt{h^{(r)}}r^i\na_i\phi\Big)\Bigg]d^3x~. \label{notmomsp}
\end{align}
The presence of the last term in \eqref{notmomsp} provides the extra term which is required to be fixed in addition to the conjugate momenta $\sqrt{h^{(r)}}\Pi^{ab}_{(r)}$. Thus, the gravitational action $\mathcal{A}$ cannot be considered as the action of the momentum space (unlike what we found in the Einstein frame). However, if we consider the gravitational action of the Jordan frame as $\mathcal{A}'$ (which has been defined in section \ref{sect2}) instead of $\mathcal{A}$, the required surface term for the well-posed action could be identified as
\begin{align}
\mathcal{A}_{sur}'=-\frac{\epsilon}{8\pi}\int_{\p\nu}\sqrt{h^{(r)}}\Big[\phi \theta^{(r)}-\frac{3}{2}r^i\na_i\phi\Big]d^3x\equiv -\frac{1}{8\pi}\int_{\nu}\mg\Big[\na_a\Big(\phi \theta^{(r)}r^a\Big)-\frac{3}{2}\square\phi\Big]d^4x~.
\end{align}
Note that this modification in the action, does not alter the expressions of the conjugate momenta and the expression of the well-posed action $\mathcal{A}_{WP}=\mathcal{A}+\mathcal{A}_{sur}=\mathcal{A}'+\mathcal{A}_{sur}'$~. Unlike $\mathcal{A}_{sur}$, $\mathcal{A}_{sur}'$ can be obtained as the integration of the product of $\Pi^{ab}_{(r)}$ and $h_{ab}^{(r)}$ \textit{i.e.} $\mathcal{A}_{sur}'=-\epsilon\int_{\p\nu}\sqrt{h^{(r)}}\Pi^{ab}_{(r)}h_{ab}^{(r)}d^3x~.$ Therefore, the modified action  $\mathcal{A}'=\mathcal{A}_{WP}-\mathcal{A}_{sur}'$  can be interpreted as the action of the momentum space. Its variation is given as follows. 
\begin{align}
\d\mathcal{A}'=\int_\nu\sqrt{-g}\Big(E_{ab}\delta g^{ab}+E_{(\phi)}\d\phi\Big)d^4x+\epsilon\int_{\p\nu}\Bigg[\sqrt{h^{(r)}}\Big(~^{(r)}D_a T^a_{(r)}+\Pi_{(\phi)}^{(r)}\d\phi\Big)+h_{ab}^{(r)}\d\Big(\sqrt{h^{(r)}}\Pi^{ab}_{(r)}\Big)\Bigg]d^3x~.
\end{align}
This agrees with the discussions provided in section \ref{sect2}, where we have found (using the bulk and surface decomposition analysis) that the gravitational action in the Jordan frame ($\mathcal{A}$) cannot be interpreted as the action of the momentum space. Hence there exists an inequivalence even at the classical level. This inequivalence can be removed if we incorporate the $\square\phi$ term in the action, as the modified action $\mathcal{A}'$ can be interpreted as the action of the momentum space. However, as we have noticed above, the $\square\phi$ term only contributes to the surface term and not on the dynamics or the conjugate momenta. Hence, it will also not contribute to the Brown-York tensor which we obtain in the following. Therefore, the following discussion has been done under the consideration of the action in the Jordan frame as $\mathcal{A}$. However, in black hole thermodynamics, we have shown earlier \cite{Bhattacharya:2018xlq} that the thermodynamic parameters depend on the surface term. Hence, in that case, the right approach was to consider the action in the Jordan frame as $\mathcal{A}'$ instead of $\mathcal{A}$ (for more discussions in this regard, please follow \cite{Bhattacharya:2020jgk,Bhattacharya:2022mnb}).

The above discussion has been presented for a particular timelike/spacelike surface. Our final goal is to obtain the well-posed action for the given manifold ($\mathcal{M}$) and to obtain the Brown-York tensor and the quasi-local parameters. We follow the same procedure as of the Einstein frame to obtain the Brown-York tensor. In the Jordan frame, the unit (spacelike) normal on $^3\mathcal{B}$ is denoted as $s_a$ and the unit (timelike) normal on $\Sigma$ is defined as $n_a$. Furthermore, the two surface $\mathcal{B}=~^3\mathcal{B}\cap \Sigma$ will be characterized by both the normals $n_a$ and $s_a$. On these surfaces, the induced metric and the extrinsic curvature are provided in the following table.
\begin{table}[H]
\begin{tabular}{ |p{3 cm}|p{3 cm}|p{3 cm}|p{3 cm}|p{3 cm}| } 
 \hline
 \textbf{Surface} & \textbf{Normal(s)} & \textbf{Induced metric} & \textbf{Extrinsic curvature} & \textbf{Trace of the extrinsic curvature} \\ 
 \hline
 ~~~~~~~~~~~$\Sigma$ & ~~~~~~~~~~~$n_a$ & $h_{ab}=g_{ab}+n_an_b$ & ~$K_{ab}=-h^i_a\na_in_b$& $K=-\na_a n^a$\\
 \hline 
 ~~~~~~~~~~$^3\mathcal{B}$ & ~~~~~~~~~~~$s_a$ & $\gamma_{ab}=g_{ab}-s_as_b$ &~$\K_{ab}=-\gamma^i_a\na_is_b$ & $\K=-\na_as^a$ \\ 
 \hline
 ~~~~~~~~~~~$\mathcal{B}$ & ~~~~~~$n_a$ and $s_a$ &$q_{ab}= g_{ab}+ n_a n_b- s_a s_b$ &~$k_{ab}=- q^i_a q^j_b\na_i s_j$ &$ k=- q^{ij}\na_i s_j=(\K_{ab}-\K \gamma_{ab}) n^a n^b$\\
 \hline
\end{tabular}
 \caption{\label{JORTABLE} Normals and extrinsic curvatures of different surfaces of the manifold in the Jordan frame}
 \end{table}
With these definitions (as provided in table \ref{JORTABLE}), we can define the well-posed gravitational action as
\begin{align}
\mathcal{A}_{tot}=\mathcal{A}-\frac{1}{8\pi}\int_{^3\mathcal{B}}\sqrt{\gamma}\phi\K d^3x+\frac{1}{8\pi}\int_{t_i}^{t_f}\sqrt{ h}\phi K d^3x~, \label{actot}
\end{align}
and the total variation of the above action \eqref{actot} yields
\begin{align}
\d\mathcal{A}_{tot}=\int_{\mathcal{M}}
\sqrt{-g}\Big[E_{ab}\delta{g^{ab}}+E_{(\phi)}\delta\phi
\Big]d^4x+\int_{^3\mathcal{B}}\sqrt{\gamma}\Big(\mathcal{D}_a\mathcal{T}^a+\Pi_{ab}\d\gamma^{ab}+\Pi_{(\phi)}\d\phi\Big)d^3x
\nonumber
\\
-\int_{t_i}^{t_f}\sqrt{h}\Big(D_a T^a+ P_{ab}\d h^{ab}+P_{(\phi)}\d\phi\Big)d^3x~.
\end{align}
Here $\mathcal{D}_a$ and $D_a$ are the covariant derivative operators compatible with $\gamma_{ab}$ and $h_{ab}$ respectively. The conjugate quantities and the total derivative terms are defined as
\begin{eqnarray}
&&\Pi_{ab}=\frac{1}{16\pi}\Big[\phi(\K\gamma_{ab}-\K_{ab})-\gamma_{ab}s^i\na_i\phi \Big]~~~~~~~~~  P_{ab}=\frac{1}{16\pi}\Big[ \phi(K h_{ab}- K_{ab})-h_{ab}n^i\na_i\phi\Big]
\no 
\\
&&\Pi_{(\phi)}=-\frac{1}{8\pi}\Big[\K+\frac{\omega}{\phi}s^i\na_i\phi\Big]~~~~~~~~~~~~~~~~~~~~~~~~P_{(\phi)}=-\frac{1}{8\pi}\Big[K+\frac{\omega}{\phi}n^i\na_i\phi\Big]
\nonumber
\\
&&\mathcal{T}^a=\frac{1}{16\pi}\phi\gamma_i^a s_j\d g^{ij}~~~~~~~~~~~~~~~~~~~~~~~~~~~~~~~~~~~ T^a=\frac{1}{16\pi}\phi h_i^a n_j \d g^{ij}      ~.
\end{eqnarray}
Thus, the surface stress tensor (or the Brown-York tensor) in the Jordan frame can be defined as
\begin{align}
T_{ab}^{(BY)}=-\frac{2}{\sqrt{\gamma}}\frac{\d\mathcal{A}_{tot}}{\d\gamma^{ab}}=-2\Pi_{ab}=\frac{1}{8\pi}\Big[\phi(\K_{ab}-\K\gamma_{ab})+\gamma_{ab}s^i\na_i\phi \Big]\label{BYEM}
\end{align}
From the above expression of the Brown-York tensor, one can obtain the expression of quasi-local energy density (or the Brown-York energy density), surface tangential momentum density and spatial stress as
\begin{eqnarray}
 &&\epsilon^{(BY)}= T_{ab}^{(BY)}  n^a n^b=\frac{1}{8\pi}[\phi k-s^i\na_i\phi]~,
\no 
\\
&& j^a= T_{bc}^{(BY)}  n^b  q^{ac}~,
\no 
\\
&& s^{ab}=q^{ac}  q^{bd} T_{cd}^{(BY)}~. \label{BYPARAJORTIME}
\end{eqnarray}
Above we have obtained the density of the quasi-local parameters from the Brown-York tensor. In the following, we make further comments where we compare the quasi-local parameters in the two frames.
\subsection{Connection of the quasi-local parameters in the two frames}
Connection of different quantities in the two frames can be obtained straightforwardly once we know how the normals in the two frames are related. The connection of the normals in the two frames can be obtained in the following manner. Consider the timelike normals $\t n_a$ and $n_a$. These normals have the following general expressions: $\t n_a=\t c\p_t$ and $ n_a=c\p_t$. However, since these are normalized to unity \textit{i.e.}, $\t n_a\t n^a=n_an^a=-1$, $\t c$ and $c$ can be fixed as $c=-1/\sqrt{-g^{00}}$ and $\t c=-1/\sqrt{-\t g^{00}}$. This, establishes the connection of the normals and other quantities in the two frames, which has been mentioned in the following table.

\begin{center}
\begin{table}[H]
\begin{tabular}{ |c|c| } 
\hline
\textbf{Quantities} & \textbf{~~~~~~~~~~Connection across the two frames~~~~~~~~~~} \\
\hline
\multirow{3}{5em}{\textbf{Normal(s)}} & $\t n_a=\sqrt{\phi} n_a ~~~~~~~~ \t s_a=\sqrt{\phi} s_a~,$ \\ 
& $\t n^a=\frac{n^a}{\sqrt{\phi}} ~~~~~~~~~~~ \t s^a=\frac{s^a}{\sqrt{\phi}}$  \\ 
\hline
\textbf{ Extrinsic curvature of $^{(3)}\mathcal{B}$} & $\tK_{ab}=\sqrt{\phi}\K_{ab}-\frac{1}{2\sqrt{\phi}}\gamma_{ab}s^i\na_i\phi$\\
\hline
\textbf{ Extrinsic curvature of $\mathcal{B}$} & $\t k_{ab}=\sqrt{\phi}k_{ab}-\frac{1}{2\sqrt{\phi}}q_{ab}s^i\na_i\phi$\\
\hline
\multirow{3}{11em}{\textbf{Trace(s) of extrinsic curvature(s)}} & $\tK=\frac{\K}{\sqrt{\phi}}-\frac{3}{2\phi^{\frac{3}{2}}}s^i\na_i\phi$ \\ 
& $\t k=\frac{k}{\sqrt{\phi}}-\frac{1}{\phi^{\frac{3}{2}}}s^i\na_i\phi$  \\ 
\hline
\textbf{ Brown-York tensor} & $\t T_{ab}^{(BY)}=\frac{T_{ab}^{(BY)}}{\sqrt{\phi}}$\\
\hline
\multirow{3}{11em}{\textbf{Quasi-local parameters}} & $\t\epsilon^{(BY)}=\phi^{-\frac{3}{2}}\epsilon^{(BY)}$ \\ 
& $ \t j^a=\phi^{-2}  j^a$  \\ 
& $\t s^{ab}=\phi^{-\frac{5}{2}}s^{ab}$  \\ 
\hline
\end{tabular}
 \caption{\label{CONNTABLE} Connection of different quantities of the two frames}
\end{table}
\end{center}

In Jordan frame, the Brown-York energy is defined as
\begin{align}
E^{(BY)}=\int_{\mathcal{B}}\sqrt{q}(\epsilon^{(BY)}-\epsilon^{(BY)}_0)d^2x,
\end{align}
where $\epsilon^{(BY)}_0$ provides the contribution of the reference spacetime. Also, $q$ is the determinant of the induced metric $q_{ab}$.
 Thus, we obtain that the Brown-York energy in the two frame are related as (which has also been found in \cite{Bose:1998yp,Cote:2019fkf})
\begin{align}
E^{(BY)}=\sqrt{\phi}\t E^{(BY)}~. \label{BYENERGYCON}
\end{align}
For the presence of a timelike Killing vector $\chi^a$ on the boundary, one can define an associated quasi-local mass \cite{Brown:1994gs}, which is defined as
\begin{align}
M^{(BY)}=\int_{\mathcal{B}}\sqrt{q} N(\epsilon^{(BY)}-\epsilon^{(BY)}_0)d^2x,
\end{align}
where $N$ is the lapse function which is related to $\chi^a=N n^a$. The lapse functions in the two frames are connected as $\t N=\sqrt{\phi} N$ (which also implies $\t\chi^a=\chi^a$). As a result, it can be shown \cite{Bose:1998yp} that the Brown-York mass is invariant in the two frames \textit{i.e.}
\begin{align}
M^{(BY)}=\t M^{(BY)}~. \label{BYMASSCON}
\end{align}
Furthermore, in the presence of a rotational Killing vector $\xi^a$, the corresponding angular momentum is obtained as 
\begin{align}
J=\int_{\mathcal{B}}\sqrt{q} j_a\xi^a d^2x.
\end{align}
Since $\t \xi^a=\xi^a$ \cite{Bhattacharya:2017pqc}, it can be proved that the angular momentum are also equivalent in the two frames \textit{i.e.}
\begin{align}
\t J^a=J^a~. \label{BYANGULARCON}
\end{align}
Although the transformation of BY mass and energy has been discussed in earlier work \cite{Bose:1998yp,Cote:2019fkf}, the conformal equivalence of the angular momenta, obtained from the BY tensor, has not been shown earlier. Furthermore, the spatial stress in the Jordan frame, as defined in eq. \eqref{BYPARAJORTIME} can be expressed as the stress tensor of a two-dimensional viscous fluid \textit{i.e.}
\begin{align}
s^{ab}=2\eta\sigma^{ab}+q^{ab}(\zeta\Theta-P)
\end{align}
where the shear tensor $\sigma^{ab}$ is the traceless part of $k^{ab}$, \textit{i.e.} $\sigma^{ab}=k^{ab}-kq^{ab}/2$~.The bulk viscosity ($\Theta$) is given as $\Theta=k$ and the expression of pressure is identified as $P=-(\phi a_i^{(n)}+\na_i\phi)s^i/8\pi$, where $a_i^{(n)}$ is given as $a_i^{(n)}=n^a\na_an_i$~. In addition, the shear viscosity coefficient ($\eta$) and the bulk viscosity coefficient $\zeta$ can be identified respectively as $\eta=\phi/16\pi$ and $\zeta=-\phi/16\pi$. Therefore, the analogy of fluid membrane can also be provided in the Jordan frame as well. However, one important remark in this regard is that the two frames are not equivalent from this fluid-gravity analogy. It is obvious that the shear- and bulk viscosity coefficients are not equivalent in the two frames. In addition, the other fluid parameters are connected in the two frames in the following manner.
\begin{align}
\sigma^{ab}=\phi^{\frac{3}{2}}\t \sigma^{ab},~~~~~~~\t\Theta=\phi^{-\frac{3}{2}}\Big(\phi\Theta-s^i\na_i\phi\Big)~~~~~\textrm{and}~~~~~\t P=\phi^{-\frac{3}{2}}\Big(P+\frac{1}{16\pi}s^i\na_i\phi \Big)~.
\end{align}
Although the fluid parameters are not equivalent in the two frames, one can show that the ratio of the shear viscosity coefficient ($\eta$) to the entropy density $s=\phi/4$) matches to the saturation value of the KSS bound and agrees to that of the Einstein frame \textit{i.e.}
\begin{align}
\frac{\eta}{s}=\frac{1}{4\pi}.
\end{align}

Furthermore, it is also noteworthy that although the original actions in the two frames are not exactly equivalent in the two frames (they are connected by the relation \eqref{ACTUAL}), the well-posed action on any timelike/spacelike surface are equivalent in the two frames \textit{i.e.}
\begin{align}
\t{\mathcal{A}}_{WP}=\t{\mathcal{A}}+\t{\mathcal{A}}_{sur}=\mathcal{A}_{WP}=\mathcal{A}+\mathcal{A}_{sur}=\mathcal{A}'+\mathcal{A}_{sur}'~. \label{AWPequivalent}
\end{align}

So far, the entire analysis is for timelike/spacelike surface. However, since the formalism is foliation dependent, the formalism mentioned above will not be straightforwardly applicable for the null surface as the null surface is degenerate. Therefore, in order to complete the picture, we have provided the same discussion for the null surface in the following section.


Before proceeding to the discussions in the following section, we summarize the important findings of this section as follows: (i) Although the usual gravitational actions in the two frames are not exactly equivalent, the well-posed actions (gravitational action+ boundary term) are exactly equivalent in the two frames.
  (ii) Again, in a different approach, it is proven that the usual gravitational action in the Jordan frame ($\mathcal{A}$) cannot be interpreted as the action of the momentum space as opposed to the modified action $\mathcal{A}'$.
  (iii) Since the dynamics of $\mathcal{A}$ and $\mathcal{A}'$ are the same, the modification in the action ($\mathcal{A}\longrightarrow \mathcal{A}'$) does not alter the expression of the Brown-York tensor and the Brown-York parameters.
  (iv) Borwn-York mass and angular momentum are conformal invariant, whereas the Brown-York energy is not invariant.
  (v) The two frames are not equivalent from the viewpoint of membrane paradigm. However, the ratio of shear viscosity to the entropy density are the same in the two frames and satisfy the celebrated result known as the KSS bound.

\section{Brown-York charges on a null-surface} \label{secnull}
In the earlier section, we have obtained the boundary term (required to define a well-posed action principle) and the Brown-York quasi-local parameters in the two frames. Also, we have obtained how the quasi-local parameters are connected under the conformal transformation. So far, the entire analysis is performed for a timelike/ spacelike surface. Since the null surface is distinctively different from a spacelike/ timelike surface, the above analysis will not be applicable for the null surface. In this section, we extend the above analysis for a null surface. Even in GR, the boundary terms and the Brown-York formalism have been developed long ago for timelike/spacelike surface. Whereas, for the null surface, the boundary term \cite{Parattu:2015gga,Parattu:2016trq,Lehner:2016vdi,Hopfmuller:2016scf,Oliveri:2019gvm,Aghapour:2018icu,Chandrasekaran:2020wwn} and the Brown-York quasi-local parameters \cite{Jafari:2019bpw,Chandrasekaran:2021hxc} have been obtained quite recently. The main reason for such difficulty is that the null surface is degenerate. Therefore, one cannot define a proper projection tensor on a null surface as opposed to a timelike or spacelike hypersurface. In the following discussion, we obtain the boundary term and the Brown-York quasi-local parameters in scalar-tensor theory for a null hypersurface. Again, for the simplicity, we start with the Einstein frame, and then the analysis in the Jordan frame will follow. Furthermore, again we do not pre-impose the boundary term to cancel the problematic terms on the boundary. We let the action principle decide the boundary term. 

\subsection{Einstein frame}
We consider, in the four-dimensional manifold ($\mathcal{M}$, $\t g_{ab}$), the null hypersurface ($\mathcal{H}$, $\t{\gamma}_{\a\b}$) is defined as, say, $\psi=$const. The main property of the null surface is that it is degenerate, which means it is possible to find the vector $\t v^{\a}$ which lies on the tangent plane of $\H$ that satisfies the condition $\t \gamma_{\a\b}\t v^{\a}=0$~. The null surface $\H$ is also characterised by the null normal $\t l_a=\t A\p_a\psi$, which is self-orthogonal \textit{i.e.} $\t l^a\t l_a=0$. For this reason, we cannot uniquely determine $\t A$. This leaves us with the major challenge of determining $\d \t l_a$ and $\d \t l^a$ as it has been discussed in literature \cite{Parattu:2015gga,Lehner:2016vdi,Hopfmuller:2016scf}. However, in keeping with the usual prescription of literature \cite{Parattu:2015gga,Jafari:2019bpw,Chandrasekaran:2021hxc}, we keep $\t A=1$. This sets $\d \t l_a=0$ (however, $\d \t l^a\neq 0$).

 Since $\t l_a$ is self-orthogonal, it is not possible to define an induced metric on $\H$. Furthermore, the null normals can be shown to satisfy the geodesic equation $\t l^a\t\na_a\t l^b=\t\kappa \t l^b$, where $\t\kappa$ is the non-affinity parameter which corresponds to the surface gravity when the null surface in consideration is a black hole horizon. As one cannot define the induced metric by $\t l^a$ alone, it was suggested by Carter \cite{Hawking:1979ig} to choose an auxiliary vector $\t k^a$, which lays out of the surface. The normalization of the null normals are considered everywhere as
\begin{align}
\t l^a\t l_a=0, ~~~~~~\t k^a\t k_a=0, ~~~~~~ \t l^a\t k_a=-1~. \label{lkdeftil}
\end{align}
We consider those variation in the metric tensor which keeps the null surface null. Therefore, the above relation \eqref{lkdeftil} will be respected by the variations. We now focus to obtain the contribution of the boundary term of the action $\d\t A$ on the null hypersurface.

In Einstein frame, the boundary contribution from the gravitational action \eqref{SE} can be obtained as (using Stoke's theorem)
\begin{align}
\t{\mathcal{B}}^{(null)}=\int_{\p\nu}\frac{\sqrt{-\t g }\t l_a}{\t A}\t\T^a d^3x=\int_{\p\nu}\frac{\sqrt{-\t g}}{\t A}\Big[\frac{1}{16\pi}\t l_a\d\t v^a-(\t l_a\t\na^a\t\phi)\d\t\phi\Big]d^3x
\end{align}
As we have discussed above, we do not have a unique way to determine $\t A$ (unlike timelike/ spacelike scenario). Therefore, using the standard prescription of the literature \cite{Parattu:2015gga,Jafari:2019bpw,Chandrasekaran:2021hxc}, from here on we set $\t A=1$.  Using \eqref{identity}, we finally obtain (for mathematical details in Einstein frame, please see the analysis in GR \cite{Parattu:2015gga,Parattu:2016trq})
\begin{align}
\tmg\t l_a\t\T^a=\frac{1}{16\pi}\Big[\p_a[\tmg\t p^a_{~b}\t l^b_{\bot}]-2\d[\tmg (\t\theta^{(l)}+\t\kappa)]\Big]+\tmg \t{\mathcal{P}}_{ab}\d\t q^{ab}
+\tmg\t{\mathcal{P}}^{(\t l)}_a\d\t l^a
\no 
\\-\tmg(\t l^a\t\na_a\t\phi)\d\t\phi~. \label{bounnulltil}
\end{align}
where 
\begin{align}
\t{\mathcal{P}}_{ab}=\frac{1}{16\pi}\Big(\t\theta_{ab}^{(l)}-(\t\theta^{(l)}+\t\kappa)\t q_{ab}\Big), ~~~~~~~~~\t{\mathcal{P}}^{(\t l)}_a=\frac{1}{8\pi}\Big((\t\theta^{(l)}+\t\kappa)\t k_a-\t k^i\tna_a\t l_i\Big)
\end{align}
and $\t\theta_{ab}^{(l)}$ and $\t\k$ are defined by the relations: $\t\theta_{ab}^{(l)}=\t q_a^i\t q_b^j\tna_i\t l_j$ and $\t l^a\tna_a\t l_b=\t\k \t l_b$~. Furthermore, $\t p^a_{~b}$ and $q^a_b$ are the two projection operators, which are defined as follows:

\begin{align}
\t p^a_{~b}=\d^a_b+\t k^a\t l_b~.
\no 
\\
\t q^a_b=\d^a_b+\t k^a\t l_b+\t l^a\t k_b~.
\end{align}

Note that $\t p^a_{~b}$ projects orthogonal to $\t l_a$, whereas $\t q^a_b$ projects orthogonal to both $\t l_a$ and $\t k_a$~. We emphasize that $\t p^a_{~b}$ is simply the projection operator and not any induced metric of $\H$. On the contrary, $\t q_{ab}$ is indeed the projection metric but not on the three-dimensional surface $\H$. Instead, it is an induced metric on the two surface, upon which both $\t l^a$ and $\t k^a$ are the normals (similar to the two surface $\mathcal{B}$ for the timelike/ spacelike surface).

Again, it can be noticed (from \eqref{bounnulltil}) that all the independent (\textit{i.e.} ten) components of the metric tensor $\t g_{ab}$ are not required to be fixed on the boundary. Instead, one has to fix three independent components of $\t q^{ab}$ ($\t q^{ab}$ has three independent components as $\t q^{ab}$ is a symmetric tensor with the constraint relations $\t q^{ab}\t l_a=0$ and $\t q^{ab}\t k_a=0$) and three independent components of $\t l^a$ (as the four vector $\t l^a$ satisfies the constraint relation $\t l^a\t l_a=0$). Thus, similar to the timelike case, the total components which are required to be fixed on the boundary are seven (one extra component of the scalar field $\t\phi$). In addition, we find the total well-posed action for a null surface can be identified as
\begin{align}
\tilde{\mathcal{A}}_{tot}|_{(null)}=\tilde{\mathcal{A}}+\frac{1}{8\pi}\int_{\H}\tmg (\t\theta^{(l)}+\t\kappa)d^3x~.
\end{align}
Obtaining the Brown-York tensor for the null surface, however, can be tricky. Since we cannot define the induced metric on the null surface, we cannot obtain the Brown-York energy-momentum tensor using the earlier definition of the timelike surface (given in eq. \eqref{BYEMTIL}). However, there can be another way to connect the above variation for the null surface with the same for the timelike one. Note that $\t\Pi_{ab}\d\t\gamma^{ab}$ (of eq. \eqref{varBYtil}) can be further decomposed as $\t\Pi_{ab}\d\t\gamma^{ab}=\t{\mathds{P}}_{ab}\d\t q^{ab}+\t{\mathds{P}}^{(\t n)}_a\d\t n^a$ (as $\t q^{ab}$, defined in Table \ref{EINTABLE}, can be written as $\t q^{ab}=\t\gamma^{ab}+\t n^a\t n^b$), which can be compared with the expression provided in \eqref{bounnulltil}. The expression of the Brown-York tensor for the timelike surface ($\t T_{ab}^{(BY)}$ which is obtained in \eqref{BYEMTIL}) can be obtained in terms of $\t{\mathds{P}}_{ab}$ and $\t{\mathds{P}}^{(\t n)}_a$ as $\t T_{b}^{a(BY)}=2 \t q^{ai}\t{\mathds{P}}_{ib}+\t n^a\t{\mathds{P}}^{(\t n)}_b$. This provides the hint of the expression of the BY tensor for a null surface in Einstein frame. In addition, while obtaining the Brown-York tensor for the null surface in GR \cite{Chandrasekaran:2021hxc}, arguments from different viewpoints also suggests that the expression of the BY tensor for a null surface has the expression $2 \t q^{ai}\t{\mathcal{P}}_{ib}+\t l^a\t{\mathcal{P}}^{(\t l)}_b$. Thus, the Brown-York tensor for the null surface in Einstein frame can be identified as
\begin{align}
\t T^a_{~~b}|_{(null)}=2 \t q^{ai}\t{\mathcal{P}}_{ib}+\t l^a\t{\mathcal{P}}^{(\t l)}_b=\frac{1}{8\pi}\Big[\t W^a_{~b}-\t p^a_{~b}\t W \Big]~,\label{BYEMNULLTIL}
\end{align}
where $\t W^a_{~b}=\t\theta^{(l)a}_{~~~b}-\t l^a\t k^i\tna_b\t l_i$ and $\t W=\t W^a_{~a}=\t\theta^{(l)}+\t\k$~. Note that the structure of the BY tensor in \eqref{BYEMNULLTIL} is similar to the timelike case as described in \eqref{BYEMTIL}, only $\tK^a_{b}$ is replaced by $\t W^a_{~b}$ and $\t\gamma^a_b$ is replaced by $\t p^a_{~b}$ in this case. However, one major difference from the timelike surface is that for null surface, the energy-momentum tensor does not appear to be symmetric. This has been noticed for Einstein's gravity as well \cite{Chandrasekaran:2021hxc,Adami:2023fbm}. The major reason for such a case could be the fact that one cannot construct a symmetric induced metric on $\H$ unlike the timelike surface. The quasi-local parameters can be identified as
\begin{eqnarray}
&&\t \epsilon^{(BY)}|_{(null)}=\t T^a_{~~b}|_{(null)} \t k_a\t l^b=\frac{\t\theta^{(l)}}{8\pi}~,
\no 
\\
&&\t j^c|_{(null)}=\t T^a_{~~b}|_{(null)} \t k_a \t q^{bc}=\frac{\t\Omega^c}{8\pi}~,
\no 
\\
&&\t s^{ab}|_{(null)}=\t q^{a}_c \t q^{bd}\t T^c_{~~d}|_{(null)}=\frac{1}{8\pi}\Big[\t\theta^{(l)ab}-\Big(\t\theta^{(l)}+\t\k\Big)\t q^{ab}\Big]~. \label{quasinulltil}
\end{eqnarray}
Here $\t\Omega^a$ is given as $\t\Omega^a=\t q^{ab}\t\omega_b$, where $\t\omega_a=\t l^i\t\na_i\t k_a$ is known as the rotation 1-form \cite{Gourgoulhon:2005ng}. Notice that although the BY tensor is not symmetric, the spatial stress on the two surface (upon which both $\t l^a$ and $\t k^a$ are normals) is a symmetric tensor. From the above quasi-local densities, defined in the above eq. \eqref{quasinulltil}, the total parameters (such as BY energy, angular momentum \textit{etc.}) can be defined as earlier (that we have defined for the timelike surface). Furthermore, the spatial stress can be obtained in the form of expression provided in \eqref{fluidtil}, where the values of shear viscosity coefficient and bulk viscosity coefficient remains unchanged (\textit{i.e.} as of the timelike surface). The expression of the shear tensor is provided as $\t\sigma^{ab}=\t\theta^{(l)ab}-\t\theta^{(l)}\t q^{ab}/2$, the expression of the bulk viscosity is provided as $\t\Theta=\t\theta^{(l)}$, and the expression of pressure is provided as $\t P=\t\kappa/8\pi$. Thus, the analogy of membrane paradigm can be provided for the null surface as well. In addition, the KSS bound will also be satisfied and the ratio of $\t\eta/s$ will be the same as of the timelike surface.
\subsection{Jordan frame}
We consider the same null hypersurface ($\H$, $\gamma_{ab}$) which is inside the four-dimensional manifold ($\mathcal{M}$, $g_{ab}$). Again, the null surface $\H$ is defined by $\psi=$const. and the null surface is degenerate. Furthermore, the normal which characterizes the null surface is defined as $l_a=A\p_a\psi$, and it is self orthogonal \textit{i.e.} $l^al_a=0$. As a result, $A$ is not uniquely determined, and we, for our convenience (and also following the practice in literature \cite{Parattu:2015gga,Jafari:2019bpw,Chandrasekaran:2021hxc}), consider $A=1$. Furthermore, the null normals satisfies the geodesic equation $l^a\na_a l^b=\kappa l^b$, where $\kappa$ is the non-affinity parameter that corresponds to the surface gravity when the null surface is black hole event horizon. In addition, following Carter's prescription, we consider the auxiliary null vector $k^a$ and the normalizations of null normals are given as
\begin{align}
l^al_a=0, ~~~~~~k^ak_a=0,~~~~~~l^ak_a=-1~. \label{lkdef}
\end{align}
Using the two null normals $l^a$ and $k^a$, one can define the two projection operators as follows
\begin{align}
p^a_{~~b}=\delta^a_b+k^al_b~,
\no 
\\
\textrm{and}~~~~~~
q^a_b=\delta^a_b+l^ak^b+k^al_b~,
\end{align}
With all the above definitions, the boundary contribution from the gravitational action \eqref{SJ} can be obtained as
\begin{align}
\B^{(null)}=\int_{\H}\sqrt{-g}l_a\T^a d^3x=\frac{1}{16\pi}\int_{\H}\mg\Big[\phi l_a\delta v^a-2l_a(\nabla_b\phi)p^{iabd}\delta g_{id}-\frac{2\omega}{\phi} (l^i\na_i\phi)\d\phi\Big]d^x~.
\end{align}
Using \eqref{identity} and following the similar steps of algebra as of the Einstein frame, one obtains
\begin{align}
\sqrt{-g}l_a\T^a=\frac{1}{16\pi}\Big[\phi\Big\{\p_a\Big(\mg p^a_{~b}\d l^b_{\bot}\Big)-2\d\Big(\mg p^a_{~b}\na_a l^b\Big)+\mg\Big(\na_al_b-g_{ab}p^c_{~d}\na_c l^d\Big)\d g^{ab}\Big\}
\no 
\\
-2\mg p^{ibad}l_a(\na_b\phi)\d g_{id}-\frac{2\o}{\phi}(l^a\na_a\phi)\d\phi\Big]
\end{align}
Again, our goal is to obtain a total surface derivative and a total variation term as expressed in eq. \eqref{schmatic}. Therefore, we bring $\phi$ inside the derivative of the first term and inside the variation of the second term. Therefore, after performing some analysis (\textit{i.e.} writing $\d g_{ab}$ in terms of $\d q^{ab}$ and $\d l^a$), we finally obtain 
\begin{align}
\sqrt{-g}l_a\T^a=\frac{1}{16\pi}\Big[ \p_a\Big(\mg\phi p^a_{~b}\d l^b_{\bot}\Big)-2\d\Big(\mg\phi p^a_{~b}\na_a l^b\Big)\Big]+\mg\mathcal{P}_{ab}\d q^{ab}+\mg \mathcal{P}^{(l)}_a\d l^a
\no 
\\
+\mg \mathcal{P}_{(\phi)}\d\phi ~, \label{BCFIN}
\end{align}
where 
\begin{eqnarray}
&&\mathcal{P}_{ab}=\frac{1}{16\pi}\Big[\phi\Big\{\theta^{(l)}_{ab}-(\theta^{(l)}+\k)q_{ab}+l_a\na_b\phi-q_{ab}l^i\na_i\phi\Big\}\Big],
\no 
\\
&& \mathcal{P}^{(l)}_a=\frac{1}{8\pi}\Big[\phi\Big\{l^i\na_a k_i+(\theta^{(l)}+\k)k_a\Big\}+k_al^i\na_i\phi\Big],
\no 
\\
&& \mathcal{P}_{(\phi)}=\frac{1}{8\pi}\Big[\theta^{(l)}+\k-\frac{\o}{\phi}l^i\na_i\phi\Big].
\end{eqnarray}
The first term of \eqref{BCFIN} can be identified as a total three-derivative term (due to $p^{\psi}_{~b}=l_ap^{a}_{~b}=0$), which can be neglected. The second term indicates the counter-term which is required to be added with the gravitational action in order to define a well-posed action principle, which is given as
\begin{align}
\mathcal{A}^{(null)}_{sur}=\frac{1}{8\pi}\int_{\H}\mg\phi(\theta^{(l)}+\k)d^3x\equiv \frac{1}{8\pi}\int_{\nu}\mg\na_a\Big(\phi(\theta^{(l)}+\k)k^a\Big)d^4x~,
\end{align}
and the well-posed action in the Jordan frame for a null surface can be identified as $\mathcal{A}_{tot}=\mathcal{A}+\mathcal{A}^{(null)}_{sur}$~. In addition, it is now obvious that all the ten independent components of the metric tensor are not required to be fixed on the boundary. Instead, we need to fix three independent components of $q^{ab}$, and three independent components of $l^a$. Thus, we need to fix seven components in total (one coming from $\phi$).

The pre-requisite of the BY formalism is a well-posed action principle. In this regard, the boundary term plays a complementary role as it makes the action well-posed. Note that the boundary term is already known for GR and for ST gravity as well (particularly for the timelike surfaces). However, there is a long-standing issue is this regard: it is that the boundary terms are, generally, pre-imposed and, are not unique. Let us take the example of GR. Although in GR, the standard boundary term in GR is considered to be the Gibbons-Hawking-York (GHY) term, there can be several other terms that can be used in substitution for the GHY term (kindly see the review \cite{Charap:1982kn}). This issue of non-uniqueness can be resolved only when the boundary term is obtained directly from the action principle itself (in the context of GR, it has been discussed in \cite{Padmanabhan:2014lwa,Parattu:2015gga}). Moreover, obtaining the boundary term from the action principle has also been shown to be consistent with the analysis of dynamical degrees of freedom corresponding to the initial value problem \cite{Padmanabhan:2014lwa,Parattu:2015gga,Parattu:2016trq,Chakraborty:2017zep,Chakraborty:2018dvi}. Therefore, we keep the same spirit of \cite{Padmanabhan:2014lwa,Parattu:2015gga,Parattu:2016trq,Chakraborty:2017zep,Chakraborty:2018dvi} and obtain the boundary terms from the action principle itself. We emphasize that the term that we finally obtain for a timelike surface is already known (rather pre-imposed in earlier cases) but, as far as our knowledge, it has not been derived with the same spirit. Furthermore, to the best of our knowledge, the boundary term that we have derived for the null surface (for ST gravity) is not known earlier and is completely a new result.

  Providing the same arguments as of the Einstein frame, we obtain the expression of the Brown-York tensor as
\begin{align}
T^a_{~b}|_{(null)}=2 q^{ai}\mathcal{P}_{ib}+l^a\mathcal{P}^{(l)}_b=\frac{1}{8\pi}\Big[\phi( W^a_{~b}-W p^a_{~b})-p^a_{~b}l^i\na_i\phi\Big]~,\label{BYEMNULL}
\end{align}
where $W^a_{~b}$ is defined as $W^a_{~b}=\theta^{(l)a}_{~~~b}-l^ak^i\na_b l_i$~. Again, the structure of the BY tensor for the null hypersurface \eqref{BYEMNULL} is similar to the timelike case \eqref{BYEM}, only $\K^a_b$ is replaced by $W^a_{~b}$ and $\gamma^a_b$ is replaced by $p^a_{~b}$ in this case. But, unlike the timelike BY tensor, the null BY tensor is not symmetric for the reasons which we have discussed during the analysis in the Einstein frame. In addition, the asymmetry has also been found for the null BY tensor in GR as well \cite{Chandrasekaran:2021hxc,Adami:2023fbm}. The quasi-local parameters can be identified as follows:
\begin{eqnarray}
&& \epsilon^{(BY)}|_{(null)}=T_{ab}|_{(null)}k^al^b=\frac{1}{8\pi}\Big[\phi\theta^{(l)}+l^i\na_i\phi \Big]~,
\no 
\\
&& j^c|_{(null)}=T_{ab}|_{(null)}k^a q^{bc}=\frac{\phi\Omega^c}{8\pi}~
\no 
\\
&&s^{ab}|_{(null)}=T_{cd}|_{(null)}q^{ac}q^{bd}=\frac{1}{8\pi}\Big[\phi\Big\{ \theta^{(l)ab}-(\theta^{(l)}+\k)q^{ab}\Big\}-q^{ab}l^i\na_i\phi\Big]~, \label{quasinulljor}
\end{eqnarray}
where $\Omega^c=q^{ac}\omega_a$ and $\omega_a=l^i\na_ik_a$ is known as the rotation 1-form. Again, the spatial stress of the two-surface $s^{ab}$ is symmetric despite the asymmetry in the BY tensor. This is because one can define the induced metric of the two surface ($q_{ab}$) which is symmetric. From the above quasi-local densities (as defined in eq. \eqref{quasinulljor}), one can define the total parameters (such as Brown-York energy, momentum \textit{etc.}) using the definitions provided for the timelike surface. Furthermore, $s^{ab}$ can be interpreted as the stress-tensor of a two-dimensional viscous fluid with the identifications $\sigma^{ab}=\theta^{(l)ab}-\theta^{(l)}q^{ab}/2$, $\Theta=\theta^{(l)}$ and $P=(\phi\kappa+l^i\na_i\phi)/8\pi$. The expressions of the shear viscosity coefficient and the bulk viscosity coefficient is the same as of the timelike surface.  Thus, the analogy of membrane paradigm can be provided for the null surface in the Jordan frame as well. Finally, we see that the ratio $\eta/s$ saturates the KSS bound \textit{i.e.} $\eta/s=1/4\pi$.

We summarize the discussion of this section as follows:
  (i) Null-hypersurface is degenerate. One cannot define an induced metric on the null surface. Hence, the usual (timelike) approach of obtaining the boundary term and quasilocal parameters do not work for the null surface.
 (ii) The structure of the Brown-York is similar to that of the timelike surface. However, unlike the timelike case, the null BY tensor is not symmetric in the two indices.
 (iii) The projection of null BY tensor on the two-surface is symmetric.
  (iv) Using the null BY formalism, one can obtain the BY energy, momentum and also can obtain the interpretation of membrane paradigm.

\section{Comparison: Timelike vs Null} \label{seccompare}
Earlier, in the timelike case, we found that the quasi-local parameters in the two frames are proportional to the same of the other frame (see table \ref{CONNTABLE}) \textit{i.e.}, they differ at most by the proportional factor of $\phi^m$ (where $m=3/2$ for energy density and so on, see table \ref{CONNTABLE}). This is an interesting result, given that each constituents of the Brown-York tensor (\textit{i.e.} the extrinsic curvature and its trace) are not proportional under conformal transformation. Furthermore, we have also obtained that the parameters like BY mass and total angular momentum are equivalent in the two frames. We can obtain these connection in the two frames mainly because, for the timelike (or spacelike) case, one can uniquely determine how the corresponding normals are connected (see table \ref{CONNTABLE}) in the two frames. But, for null surface, there is no unique way to determine how the null normals are connected in the two frames. We can consistently connect the null vectors in the two frames in the following arbitrary manner:
\begin{eqnarray}
&&\t l_a=\phi^p l_a,~~~~~~~~~~~~~\t l^a=\phi^{p-1}l^a,
\no 
\\
&&\t k_a=\phi^{-p+1}k_a,~~~~~~~~\t k^a=\phi^{-p}k^a~, \label{lnkarbitrarynull}
\end{eqnarray}
where $p$ can be arbitrary. In spite of the arbitrariness, the above connection is consistent with the normalization conditions \eqref{lkdeftil} and \eqref{lkdef}. This arbitrariness is not applicable for, say, a timelike normal (\textit{i.e.} one cannot define $\t n_a=\phi^p n_a$ where $p$ is arbitrary; after all it has to satisfy $\t n^a\t n_a=n^an_a=-1$, which uniquely determines $p$). Since we cannot uniquely determine $p$ for a null surface, we cannot compare how the quasilocal parameters in the two frames are connected to the same of another frame. In addition, if we fix $p$ as per our choice, say we fix $p=0$ or $p=1$ for the simplicity in calculation, we find that the quasi-local parameters in the two frames are no longer proportional to the same of the other frame.

However, there exists a way out to get rid of this issue, which comes from the study of the boundary terms. As we discussed earlier, the gravitational action are not exactly equivalent in the two frames due to eq. \eqref{ACTUAL}. But, for timelike/spacelike surface, we earlier obtained in eq. \eqref{AWPequivalent} that the well-posed action (\textit{i.e.} the gravitational action along with the boundary term) are equivalent. This was true even if we consider the action of the Jordan frame as $\mathcal{A}'$ instead of $\mathcal{A}$. Let us assume that this equivalence of the well-posed action will be valid for the null-surface as well. For the arbitrary relation among the null vectors \eqref{lnkarbitrarynull}, we can obtain the following connections for the extrinsic curvature and the surface gravity:

\begin{eqnarray}
&&\t\theta_{ab}^{(l)}=\phi^p\Big[\theta_{ab}^{(l)}+\frac{q_{ab}}{2\phi}l^i\na_i\phi\Big]~,
\no 
\\
&&\t\theta^{(l)}=\phi^{p-2}\Big[\phi \theta^{(l)}+l^i\na_i\phi\Big]~,
\no 
\\
&& \t\k=\phi^{p-2}\Big[\phi\k+p~ l^i\na_i\phi\Big]~.
\end{eqnarray}

This implies that the surface term of the Einstein frame is related to that of the Jordan frame in the following manner.
\begin{align}
\p_a\Big[\tmg\Big(\t\theta^{(l)}+\t\k\Big)\t k^a\Big]=\p_a\Big[\mg\Big(\phi\Big(\theta^{(l)}+\k\Big)+(p+1)l^i\na_i\phi\Big)k^a\Big]~.
\end{align}
If we consider that the well-posed action in the two frames are equivalent in the two frames, it fixes $p$ as $p=1/2$. Interestingly, for this choice, we also obtain that the quasi-local parameter density are related in the same way as of the timelike case, \textit{i.e.}

\begin{align}
\epsilon^{(BY)}|_{(null)}=\phi^{\frac{3}{2}}\t\epsilon^{(BY)}|_{(null)}~,
\no 
\\
 j^a|_{(null)}=\phi^2 \t j^a|_{(null)}~,
\no 
\\
s^{ab}|_{(null)}=\phi^{\frac{5}{2}}\t s^{ab}|_{(null)}~. \label{connectionnull}
\end{align}
For this choice one can also obtain that the BY mass and the angular momentum are equivalent in the two frames and the total BY energy are related in the same way as that of the timelike case. In other words, the equations \eqref{BYENERGYCON}, \eqref{BYMASSCON} and \eqref{BYANGULARCON} will be valid for the null surface as well. 

The arbitrariness in the connection of null normals (as described by Eq. \eqref{lnkarbitrarynull}) lies in the property of the null surface itself (owing to its degeneracy). In fact, our analysis helps us to resolve this arbitrariness as we find that one can explicitly show the connection of the null normals and, thereby, the BY parameters if one claims that the well-posed action (gravitational action along with the boundary term) is equivalent. We have proved this claim for timelike/spacelike surfaces and, one can expect it to be valid for the null surface as well. After all, the equations of motions, which arise from the gravitational actions are equivalent and the problematic parts are cancelled by the boundary terms. Therefore, the well-posed actions, which are equivalent in timelike/spacelike surface, are expected to be the same for the null surface as well.

Note that the null formalism (as presented in the present section and in previous section) simply upholds the conclusions of the timelike surface (as described in section \ref{sectimelike}). Since the null normals are related arbitrarily (due to the degeneracy in the tangent plane of the null surface), one cannot straightforwardly obtain the conformal connections between the BY parameters. However, if we claim that the well-posed action, \textit{i.e.} the gravitational action along with the boundary term, is equivalent (which is the case in the timelike/spacelike case), one can obtain the connection between the null normals in the two frames and show how the BY parameters are connected in the two frames. In this case as well, we find that the BY parameters which appears in BH thermodynamics (such as the mass and angular momentum) are equivalent. On the other hand, the parameters, which are related to fluid-gravity correspondence (such as bulk and shear viscosity coefficient, pressure \textit{etc.}) are not equivalent in the two frames. Thus, the analysis in timelike surface and in null surface, provide us with the same conclusions.

We summarize the discussion of the present section as follows: (i) Unlike the timelike normals, the connection of null normals of the two frames are not uniquely determined. Therefore, one cannot uniquely determine how the BY tensor and the parameters are connected. (ii) If one assumes that the well-posed action is conformally equivalent, it leads to a specific choice on the connection of the null normals. For this choice, one can again establish the same connection relations among the parameters of the two frames.

\section{Conclusions} \label{secconcl}
The ``local'' definition of mass-energy has been the subject of intense research for a long time. However, it has not been possible so far to arrive to an unanimous conclusion. In the absence of a local definition of mass-energy, the Brown-York formalism provides a powerful way to define quasi-local parameters like mass, angular momentum, spatial stress, \textit{etc.} which are important both from the perspectives of black hole thermodynamics as well as fluid-gravity correspondence. So far, people only have studied the properties of BY mass and energy under the conformal transformation. Therefore, in order to obtain a thorough understanding, we have studied the conformal connection of all quasi-local parameters, which are provided by the Brown-York formalism. Furthermore, the important pre-requisite of Brown-York energy is that the action principle must be well-posed. However, the gravitational actions of ST gravity are not well-posed ones like the Einstein-Hilbert action. As a result, one has to incorporate proper boundary term in order to make the action principle well-defined and to obtain the Brown-York quasi-local parameters. For scalar-tensor theory, the boundary term required for the action principle on a timelike/spacelike surface is pre-imposed \textit{i.e.} not obtained in a consistent manner. On the other hand it has been argued recently that the action principle should suggest us what boundary term should be added. Therefore, we have performed our analysis in such spirit. Moreover, the boundary term for a null surface and the null Brown-York formalism has not been obtained so far in the context of ST gravity. To fill this gap, we have provided a complete analysis on the null boundary term and the null BY quasi-local parameters.

Here, we have firstly outlined the properties of the gravitational actions of the ST gravity in the two frames briefly, which has been obtained in the works of one of the authors \cite{Bhattacharya:2017pqc,Bhattacharya:2022mnb}. We have shown that the two frames are not exactly equivalent at the action level itself. Instead the gravitational actions in the two frames differ by a total derivative ($\square\phi$) term which is usually neglected. Moreover, we have also shown that the (gravitational) action in the Einstein frame can be regarded as the action of the momentum space and the well-known holographic relation can be obtained. But, the same (holographic) relation cannot be obtained for the gravitational action in the Jordan frame and, therefore, the action cannot be interpreted as the action of the momentum space. This shows that there lies an in-built inequivalence of the two frame when we neglect the $\square\phi$ term. When this term is accounted, the inequivalence in the two frames can be removed. Later, we have focused on obtaining the BY tensor and the boundary term from the single analysis of action principle. Firstly, we have made the analysis for the timelike/spacelike surface. From the variation of the gravitational action we obtain the boundary term which is required to be added with the gravitational action, the variables which are required to be fixed and the conjugate quantities corresponding to those variables. Once these quantities are obtained, we obtained the Brown-York tensor and the corresponding quasi-local parameters. Later we have repeated the same analysis for a more non-trivial case \textit{i.e.} for the null surface and obtained the same quantities.

Our analysis shows that the quantities which are related to the black hole thermodynamics (like mass, angular momentum \textit{etc.}) are conformally equivalent. However, the quantities which are related to the fluid-gravity connection (such as bulk and shear viscosity coefficient, pressure \textit{etc.}) are not conformally invariant. Also, both the analysis, which are individually performed for timelike/spacelike surface (in section \ref{sectimelike}) as well as for null surface (in section \ref{secnull} and \ref{seccompare}) yields the same conclusions (\textit{i.e.} the parameters related to black hole thermodynamics are conformally equivalent and the parameters related to the fluid-gravity analogy are inequivalent). In the recent paper by one of the authors \cite{Bhattacharya:2020wdl}, it has been found that one can have both equivalent and inequivalent pictures for fluid-gravity correspondence in ST theory while obtaining the Damour-Navier-Stokes equation in both the frames. In order to identify which of these pictures (inequivalent or in-equivalent) are more appropriate, one has to examine from other perspectives. From the viewpoint of BY formalism (as presented in this paper), we find that the fluid parameters are not equivalent and their expressions are the same as of the in-equivalent picture of the earlier work \cite{Bhattacharya:2020wdl}. This is how the present analysis favours the in-equivalent viewpoint of \cite{Bhattacharya:2020wdl}.
 Also, we mention that, although our analysis has been performed for the scalar-tensor theory, it will be valid for $f(R)$ \cite{Sotiriou:2008rp,DeFelice:2010aj} theory as well where $\phi$ will be replaced by $f'(R)=\p f(R)/\p R$.

The main goal of this paper is to provide and all-around perspective regarding the BY formalism in ST gravity. Moreover, the null BY formalism and the null boundary term is a recent development in GR. Therefore, in the present work, we have discussed BY formalism and the boundary term (which plays a complementary role in the study of BY formalism) in ST gravity both for the timelike as well as for the null surfaces. For the timelike surface, the boundary term is known. But, we have derived it from the action principle itself to be on par with the recent arguments regarding uniqueness and consistency with the analysis of the degrees of freedom. Our analysis provides a mixed view regarding the (in)equivalence of the two frames. While it supports the equivalence of the parameters which appear in BH thermodynamics, but it shows that the parameters which are related to fluid-gravity analogy, are not equivalent and favours the inequivalent picture of \cite{Bhattacharya:2020wdl}. The analysis in the null surface also supports this finding. Although apparently the analysis in the null surface seems to be inconclusive regarding the (in)equivalence, when we claim that the well-posed action is invariant under the conformal transformation (as is the case in timelike surface), it it yields the same conclusions as of the analysis in the timelike surface.

Our present analysis suggests that the fluid-gravity analogy in its current form shows to be inequivalent under the conformal transformation. Therefore, more investigations is solicited in this direction for better understanding \cite{INPREP}.
 This paper provides a robust analysis in the context of boundary term and the BY quasilocal parameters and we hope it will be a significant contribution in the understanding of ST gravity and overall behaviour of gravitational physics under the conformal transformation.

\section*{Acknowledgement} This work is supported by the JSPS KAKENHI Grant (Number: 23KF0008) and the research of K. Bamba is supported in part by the JSPS KAKENHI Grant (Number: JP21K03547).
\appendix
\section{Obtaining Eq. \eqref{identity}} \label{proofidentity}
Since $\d v^a=2P^{ibad}\na_b\d g_{id}$, we obtain
\begin{eqnarray}
&&X_a\d v^a=-X_a\na_b\d g^{ab}+X^a g_{ij}\na_a\d g^{ij}
\no 
\\
&&~~~~~~~~=-\na_b\Big(X_a\d g^{ab}\Big)+(\na_aX_b)\d g^{ab}+X^a g_{ij}\na_a\d g^{ij}~. \label{firststepa1}
\end{eqnarray}
Now, we need to compute the last term of \eqref{firststepa1}, which can be simplified as $X^a g_{ij}\na_a\d g^{ij}=X^a\p_a(g_{ij}\d g^{ij})$. Also, one can obtain 
\begin{eqnarray}
&&\p_a(g_{ij}\d g^{ij})=-\frac{2}{\mg}\p_a\Big(\d\mg\Big)-\frac{1}{\mg}g_{ij}\d g^{ij}\p_a\Big(\mg\Big)
\no 
\\
&&~~~~~~~~~~~~~~=-\frac{2}{\mg}\p_a\Big(\d\mg\Big)+\frac{2}{(\mg)^2}\d(\mg)\p_a\Big(\mg\Big)~.
\end{eqnarray}
Therefore, 
\begin{eqnarray}
&&X^a g_{ij}\na_a\d g^{ij}=-\frac{2}{\mg}X^a\p_a\Big(\d\mg\Big)+\frac{2}{(\mg)^2}\d(\mg)X^a\p_a\Big(\mg\Big)
\no 
\\
&&~~~~~~~~~~~~~~~~=-\frac{2}{\mg}\p_a\Big(X^a\d\mg\Big)+\frac{2}{\mg}\d\mg\p_aX^a+\frac{2}{(\mg)^2}\d(\mg)X^a\p_a\Big(\mg\Big)
\no 
\\
&&~~~~~~~~~~~~~~~~=-\frac{2}{\mg}\p_a\Big(X^a\d\mg\Big)+\frac{2}{\mg}\d(\mg)\na_aX^a
\no 
\\
&&~~~~~~~~~~~~~~~~=-\frac{2}{\mg}\d\Big(\p_a(\mg X^a)\Big)+\frac{2}{\mg}\p_a\Big(\mg\d X^a\Big)+\frac{2}{(\mg)^2}\d(\mg)\na_aX^a
\no 
\\
&&~~~~~~~~~~~~~~~~=-2\d\Big[\frac{1}{\mg}\Big(\p_a(\mg X^a)\Big)\Big]+2\d\Big(\frac{1}{\mg}\Big)\p_a(\mg X^a)+2\na_a(\d X^a)+\frac{2}{(\mg)^2}\d(\mg)\na_aX^a
\no 
\\
&&~~~~~~~~~~~~~~~~=-2\d\Big(\na_aX^a\Big)+\na_a\Big(\d X^a+g^{ab}\d X_b+X_b\d g^{ab}\Big)~. \label{secondstepa1}
\end{eqnarray}
Substituting Eq.\eqref{secondstepa1} in Eq. \eqref{firststepa1}, we obtain Eq. \eqref{identity}.

\section{Obtaining Eq. \eqref{boufin}} \label{appen2}
Using \eqref{identity}, the first term of \eqref{btotjor} is given as
\begin{align}
\sqrt{h^{(r)}}\phi r_a\delta v^a=\sqrt{h^{(r)}}\phi\na_a r^a_{\bot}-\sqrt{h^{(r)}}\phi\delta(2\na_a r^a)+\sqrt{h^{(r)}}\phi\na_ar_b\d g^{ab}~. \label{b1jor}
\end{align}
In the Jordan frame, the extrinsic curvature of the surface (upon which $r_a$ is the unit normal) can be defined as
\begin{align}
\theta^{(r)}_{ab}=- h_a^{(r)i}h_b^{(r)j}\na_i r_j=- h_a^{( r)i}\na_i r_b
\no 
\\
=-\na_a r_b+\epsilon  r_a  a^{(r)}_b~, \label{extrin}
\end{align}
where $a^{(r)}_i=r^a\na_a r_i$ and the trace of the extrinsic curvature is given as 
\begin{align}
\theta^{(r)}=-\na_a r^a~. \label{extrintrace}
\end{align}
In addition, the covariant derivative operator, which is adapted to the surface, can be defined as
\begin{align}
^{(r)}D_aA_b=h_a^{(r)i}h_b^{(r)j}\na_i A_j~,\label{threederivative}
\end{align}
Replacing \eqref{extrin}, \eqref{extrintrace} and \eqref{threederivative} in \eqref{b1jor}, one can obtain 
\begin{align}
\sqrt{h^{(r)}}\phi r_a\delta v^a=\sqrt{h^{(r)}}~^{(r)}D_a\Big(\phi h^{(r)a}_i r_j\d g^{ij}\Big)+2\delta\Big(\sqrt{h^{(r)}}\phi \theta^{(r)}\Big)-\sqrt{h^{(r)}}h^{(r)a}_{i}r_j\d g^{ij}\na_a\phi
\no 
\\
+ \sqrt{h^{(r)}}\phi \Big(-\theta^{(r)}_{ab}+\theta^{(r)}h^{(r)}_{ab}\Big)\d h^{(r)ab}-2\sqrt{h^{(r)}}\theta^{(r)}\d\phi~. \label{firtersim}
\end{align}
Thus, the first term of \eqref{btotjor} is obtained above in \eqref{firtersim}.
We replace \eqref{firtersim} in \eqref{btotjor} and finally obtain \eqref{boufin}.


\end{document}